\documentclass[graybox]{svmult}

\usepackage{mathptmx}       % selects Times Roman as basic font
\usepackage{helvet}         % selects Helvetica as sans-serif font
\usepackage{courier}        % selects Courier as typewriter font
\usepackage{type1cm}        % activate if the above 3 fonts are
\usepackage{color}
\usepackage{makeidx}         % allows index generation
\usepackage{graphicx}        % standard LaTeX graphics tool
\usepackage{multicol}        % used for the two-column index
\usepackage[bottom]{footmisc}% places footnotes at page bottom

\makeindex             % used for the subject index
\begin{document}

\title*{Cellular Automata Applications in Shortest Path Problem}
% Use \titlerunning{Short Title} for an abbreviated version of
% your contribution title if the original one is too long
\author{Michail-Antisthenis I. Tsompanas, Nikolaos I. Dourvas, Konstantinos Ioannidis, Georgios Ch. Sirakoulis, Rolf Hoffmann, Andrew Adamatzky}
 \authorrunning{Tsompanas, Dourvas, Ioannidis, Sirakoulis, Hoffmann, Adamatzky} 
\institute{Michail-Antisthenis I. Tsompanas \and Andrew Adamatzky \at University of the West of England, Bristol, BS16 1QY, United Kingdom, \email{Antisthenis.Tsompanas@uwe.ac.uk , Andrew.Adamatzky@uwe.ac.uk}
\and Nikolaos I. Dourvas \and Georgios Ch. Sirakoulis \at Laboratory of Electronics, Department of
Electrical and Computer Engineering, Democritus University of Thrace, Xanthi, GR67100, Greece
\email{ndourvas@ee.duth.gr, gsirak@ee.duth.gr}
\and Konstantinos Ioannidis \at Information Technologies Institute Centre for Research and Technology Hellas, 6th km Charilaou-Thermi Road 57001, Thermi, Thessaloniki, Greece, \email{kioannid@iti.gr}
\and Rolf Hoffmann \at Department of Computer Science, Technical University of Darmstadt, Hochschulstraße 10 D-64289 Darmstadt, Germany, \email{hoffmann@rbg.informatik.tu-darmstadt.de}
}
%
% Use the package "url.sty" to avoid
% problems with special characters
% used in your e-mail or web address
%

\maketitle

\begin{abstract}
{
\emph{Cellular Automata (CAs)} are computational models that can capture the essential features of systems in which global behavior emerges from the collective effect of simple components, which interact locally. During the last decades, CAs have been extensively used for mimicking several natural processes and systems to find fine solutions in many complex hard to solve computer science and engineering problems. Among them, the \emph{shortest path problem} is one of the most pronounced and highly studied problems that scientists have been trying to tackle by using a plethora of methodologies and even unconventional approaches. The proposed solutions are mainly justified by their ability to provide a correct solution in a better time complexity than the renowned Dijkstra's algorithm. Although there is a wide variety regarding the algorithmic complexity of the algorithms suggested, spanning from simplistic graph traversal algorithms to complex nature inspired and bio-mimicking algorithms, in this chapter we focus on the successful application of CAs to shortest path problem as found in various diverse disciplines like computer science, swarm robotics, computer networks, decision science and biomimicking of biological organisms' behaviour. In particular, an introduction on the first CA-based algorithm tackling the shortest path problem is provided in detail. After the short presentation of shortest path algorithms arriving from the relaxization of the CAs principles, the application of the CA-based shortest path definition on the coordinated motion of swarm robotics is also introduced. Moreover, the CA based application of shortest path finding in computer networks is presented in brief. Finally, a CA that models exactly the behavior of a biological organism, namely the Physarum's behavior, finding the minimum-length path between two points in a labyrinth is given. The CA-based model results are found in very good agreement with the computation results produced by the in-vivo experiments especially when combined with truly parallel implementations of this CA in a Field Programmable Gate Array (FPGA) and on a Graphical Processing Unit (GPU). The presented implementations succeed to take advantage of the CA's inherit parallelism and significantly improve the performance of the CA algorithm when compared with software in terms of computational speed and power consumption.}% \cite{Dijkstra1959}
\end{abstract}

%\end{frontmatter}

%\linenumbers

\section{Introduction}
\label{sec:intro}

The shortest path problem has always been a hot topic in the study of graph theory, because of its wide application field, extending from operational research to the disciplines of geography, automatic control, computer science and traffic. According to its concrete applications, scholars in relevant fields have presented many algorithms, but most of them are solely improvements \cite{Johnson} based on Dijkstra's algorithm \cite{Dijkstra}. Shortest path problems can be solved in polynomial time by one of the many shortest path algorithms, such as Dijkstra \cite{Dijkstra} and Floyd-Warshall \cite{Floyd,Warshall}, provided that edge lengths are deterministic, i.e. every feasible probability distribution, out of a given set, over all possible successor nodes assigns probability one to a single successor. On the other hand, Cellular automata (CAs) \index{cellular automata} are models of physical systems, where space and time are discrete and interactions are local \cite{Neumann}. Prior and more recent works proved that CAs are very effective in simulating physical systems and solving scientific problems, because they can capture the essential features of systems where global behavior arises from the collective effect of simple components, which interact locally \cite{adamatzky1994,acri2012,Sirakoulis,Was,Samira}. Furthermore, they can easily handle complex boundary and initial conditions, inhomogeneities and anisotropies \cite{Tsompanas,Dourvas}. The last decades, a wide variety of CA applications have been proposed on several scientific fields, such as simulation of physical systems, biological modeling involving models for self-reproduction, biological structures, image processing, semiconductor fabrication processes, crowd evacuation, computer networks and quantum CAs \cite{Sirakoulis99,Sirakoulis99a,Mardiris,Georgoudas2010,Nalpa,Konsta,Tsompanas15,Tsiftsis,Giitsidis,Kechaidou}. These problems are described in terms of CAs, spatially by an 1-d, 2-d or 3-d array of cells and a local rule, which is usually an arbitrary function that defines the new state(s) of its CA cell depending on the states of its neighbors. The CA cells can work in fully synchronous and parallel manner updating their own state. It is clear that the CA approach can be considered consistent with the modern notion of unified space time, where, in computer science, space corresponds to memory and time to processing unit. In analogy, in CA, memory (cell state) and processing unit (local rule) are inseparably related to a CA cell \cite{Ntinas,TsiftsisFPGA}. Taking all the above into consideration, there is no surprise that CA have been also able to deal successfully with the shortest path problem providing coherent and computationally efficient solutions in a number of various scientific applications and fields as shown later in this chapter. In what follows, we will focus in some of the most pronounced CA based applications that present different confrontations and corresponding solutions concerning the shortest path problem.% when it is depicted in the CA grid.

\section{The first Cellular Automata approach in Shortest Path Problem}
\label{sec:adamatzky}

The first CA algorithm tackling the shortest path problem was proposed by Adamatzky \cite{adamatzky1996computation} although, it can be claimed that the famous Lee algorithm \cite{lee} could be considered as the first CA alike approach (see the CA algorithm in Section \ref{CALee} for unweighted cells). However, the CA algorithm proposed by Adamatzky is mainly studying a weighted graph, which is also oriented. The three most common variations of the problem, namely single source shortest path (S$^3$P), all pairs shortest path (APSP) and single source single destination shortest path (S$^3$DSP), were all faced by the proposed CA algorithm. The main aim of this work was to tackle in a parallel way the S$^3$P and APSP variants by implementing a CA with adjustable neighborhood radius. Solving a shortest path problem or identifying the most direct path in a network among two vertices (i.e. $x$ and $y$) was approximated using a CA, by plotting the under study graph onto a rectangular mesh, where cells $x$ and $y$ represent the respective vertices. The proposed solution is reached when an excitation wave with starting point cell $x$, diffuses towards all directions and arrives at cell $y$.

CAs have simulated living processes, neural networks, cellular and animal populations, molecular liquids, membranes and excitable reaction-diffusion media, because they are the most material, perceptible and practical models \cite{adamatzky1996computation}. The main feature of excitation in a medium is that signals can be propagated undamped over a long distance and the speed of wave propagation can be variable. In Adamatzky's work the speed of a wave is proportional to the weight of the edge connection between node $x$ and node $y$. At the beginning the given graph is mapped onto the cellular array of CA. Source vertex $x$ and destination vertex $y$ of the graph are corresponding to cells $x$ and $y$, respectively. The $x$ cell is excited and the wave propagates in all directions around the lattice, modifying the states of cells. Computation is assumed finished when the wave reaches destination $y$. 

The definition of a CA is a $d$-dimensional lattice $L$ of $n$ cells, cell states $Q$, neighborhood function $u$ and transition function $f$. For every cell $x$ a neighborhood function assigns a group of the closest cells $u(x)$. The local transition function $f$ maps a set of neighborhood states into the set $Q$ of cells states. Using all the above characteristics, the next state of cell $x$ will be defined as the state of its neighborhood in the previous time step and the rule of the transition function in the following way: $x^{t+1}=f(u(x^t)) $. Evolution of CA with initial configuration $c$ is a series of transitions like: $c^0 \rightarrow c^1 \rightarrow c^2 \rightarrow ...c^t \rightarrow c^{t+1} \rightarrow ...$. In this (ours) work \cite{adamatzky1996computation} a pointer $p_x$ and a vector $w_{xy}$ are used, which correspond to the direction from which the wave has propagated at the previous time step and the weight of the edge between nodes $x$ and $y$, respectively. The set $Q$ can take one of the following elements: $Q=\{+,\#,\bullet,0,1,2,...,\nu\}$. Pointer $p_x$ takes values from a finite nonempty set $Y=\{1,2,\ldots, k,\lambda\}$ where $\lambda$ can be considered as the initial value of $p_x$. Every element $w_{xy}$ of vector $w_x$, $y \in u(x)$, can be in one of the states of set $W=\{\infty,0,...,\nu\} $. When vertices (nodes) $x$ and $y$ are not connected with one another by an edge oriented from $x$ to $y$ then $w_{xy}=\infty$ .

Let $G=\langle V,E\rangle$ be an oriented graph of $n$ vertices arranged on the $d$-dimensional discrete lattice, every vertex $\upsilon\in V$ of which is connected with no more than $k$ neighboring vertices $\upsilon_1,\upsilon_2,...\upsilon_k$ by input edges $\upsilon_1\upsilon,\upsilon_2\upsilon,...,\upsilon_k\upsilon \in E$ of weights $w(\upsilon_1\upsilon), w(\upsilon_2\upsilon),...,w(\upsilon_k\upsilon) \in \{0,1,...\nu,\infty \}$. If some pair $\upsilon'\upsilon'' \in V$, $w(\upsilon'\upsilon'')=\infty$ is written when there is no edge $\upsilon'\upsilon''$ in set $E$ (i.e. there is no edge between those vertices). There is a principal feature that graph $G$ must have in order to be successfully mapped onto a cellular lattice: $k<n$. Let $p=(\upsilon_0,...,\upsilon_m)$ be a shortest path from vertex $\upsilon_0$ to vertex $\upsilon_m$ of graph $G$ and $l(p)$ be a length of $p$. Then we have that $l(p)=min\{l(p'):p'=\{\upsilon_0,...,\upsilon_m\}\}$.

First, the single source, single destination shortest path ($S^3DSP$) is considered. This can be used in $S^3P$ and $APSP$ in a similar way. In the $S^3DSP$ computation, at the beggining $x^0_s=+$, where $t=0$ is assumed and $x_s$ and $x_d$ are the source and destination nodes while $+$ is the wave of excitation. The computation stops when cell $x$ passes in state $\#$ or every cell of $L$ is in state $\#$ or $\bullet$ ($\bullet$ is a quiescent-like state). The second constraint is used to stop computation when there is no path from $x_s$ to $x_d$. The virtual wave is moving in cellular lattice. The wave of states $+$ runs in all directions around the lattice from cell $x_s$ until it is in $x_d$, or passes all the cells of $L$. The pointer $p_x$ has an initial state $\lambda$ as mentioned before. When the cell $x$ takes the $+$ state then $p_x$ changes its initial value and takes one from set $\{1,2,\ldots,k\}$, which is the index of the neighbor from where the $+$ state has come from. In this way, the final path can be extracted easily from the final configuration of the CA by back-tracking over the pointers from cell $x_d$ toward cell $x_s$. The transition function $f$ works with the neighborhood function $u(x)$ and weights $w_{xy}$ in the following way. Assuming that cell $x^t=\bullet$ and some of its neighbors from $u(x)$ are in state $+$ at time $t$. Cell $x^t$ finds the neighbor $y^t=+$ that has the minimum weight value and $x$ takes this value. Starting from state $w_{xy}$, cell $x$ jumps in state 0, decreasing its current state on unit step at every time step $x^{t+1}=x^t-1$. Cell $x^t$ will take the state $+$ when $x^t=0$ or there exists a neighbor $y^t=+$ and the weight of the edge between $x^t$ and $y^t$ is the minimum and $w^t_{xy}<x^t$. The transition from state 0 to state $+$ and from state $+$ to state $\#$ is occurring unconditionally. 

When the simulation starts, pointers of all cells are in state $\lambda$. However, if cell $x$ changes its state to $w_{xy}$, then for some neighbor $y_j$ pointer $p_x$ saves the index $j$ of this neighbor. During the decreasing of $w_{xy}$ down to 0, pointer $p_x$ can be modified when there is the condition $\forall y \in u(x):y^t=+ \wedge w_{xy}=min\{w_{xy'}:y'\in u(x) \wedge y'^t=+\}$ and $w^t_{xy}<x^t$. The state of pointer becomes constant after cell $x$ departs from state 0. Concluding all these rules, $x^{t+1}$ can take the following states in every case:

\begin{itemize}
	\item $\#$, when $x^t$ is in states $\#$ or $+$.
	\item $+$, when $x^t=0$.
	\item $\bullet$, when $x^t=\bullet$ and there in no neighbor $y^t=+$.
	\item $w_{xy}$, when $x^t=\bullet$ or $x^t>0$ and there is at least one neighbor with $y^t=+$ and the weight of the edge between $x^t$ and $y^t$ is the minimum of other edges in the neighborhood.
	\item $x^t-1$, when $x^t>0$ and there is no other neighbor $y^t=+$ that has a weight in the edge between $x^t$ and $y^t$ in order that $w_{xy}<x^t$.
\end{itemize}

\begin{figure}[htbp]
	\centering
	\includegraphics[scale=0.4]{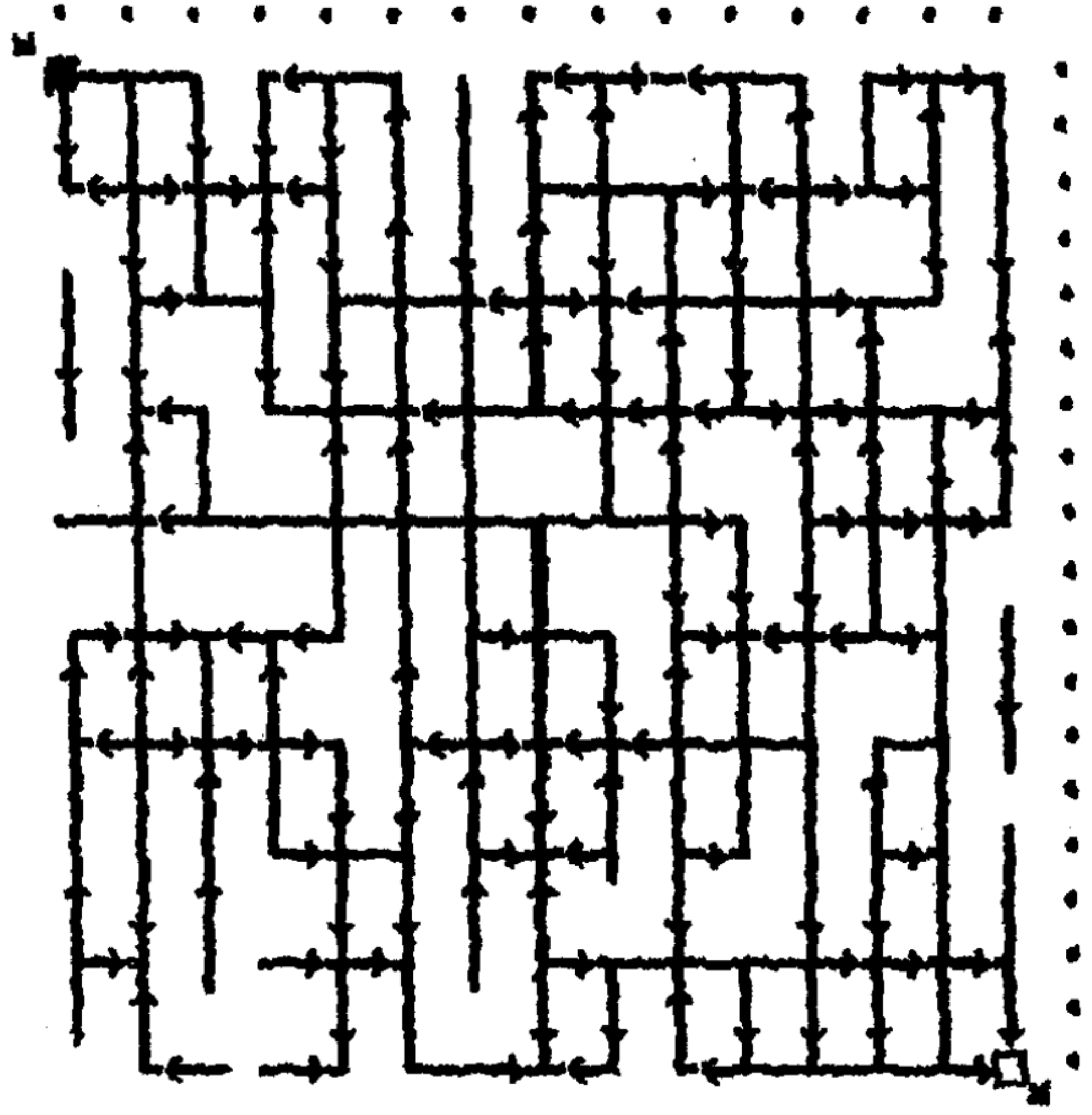}
	\caption{The example of oriented graph $G$: arrows indicate orientation of edges; and intersection of two or more straight lines corresponds to a vertex of $G$; black and empty boxes are source vertex and destination vertex, respectively (adopted from \cite{adamatzky1996computation}).}
	\label{fig:AndyGraph}
\end{figure}

\begin{figure}[htbp]
	\centering
	\includegraphics[scale=0.85]{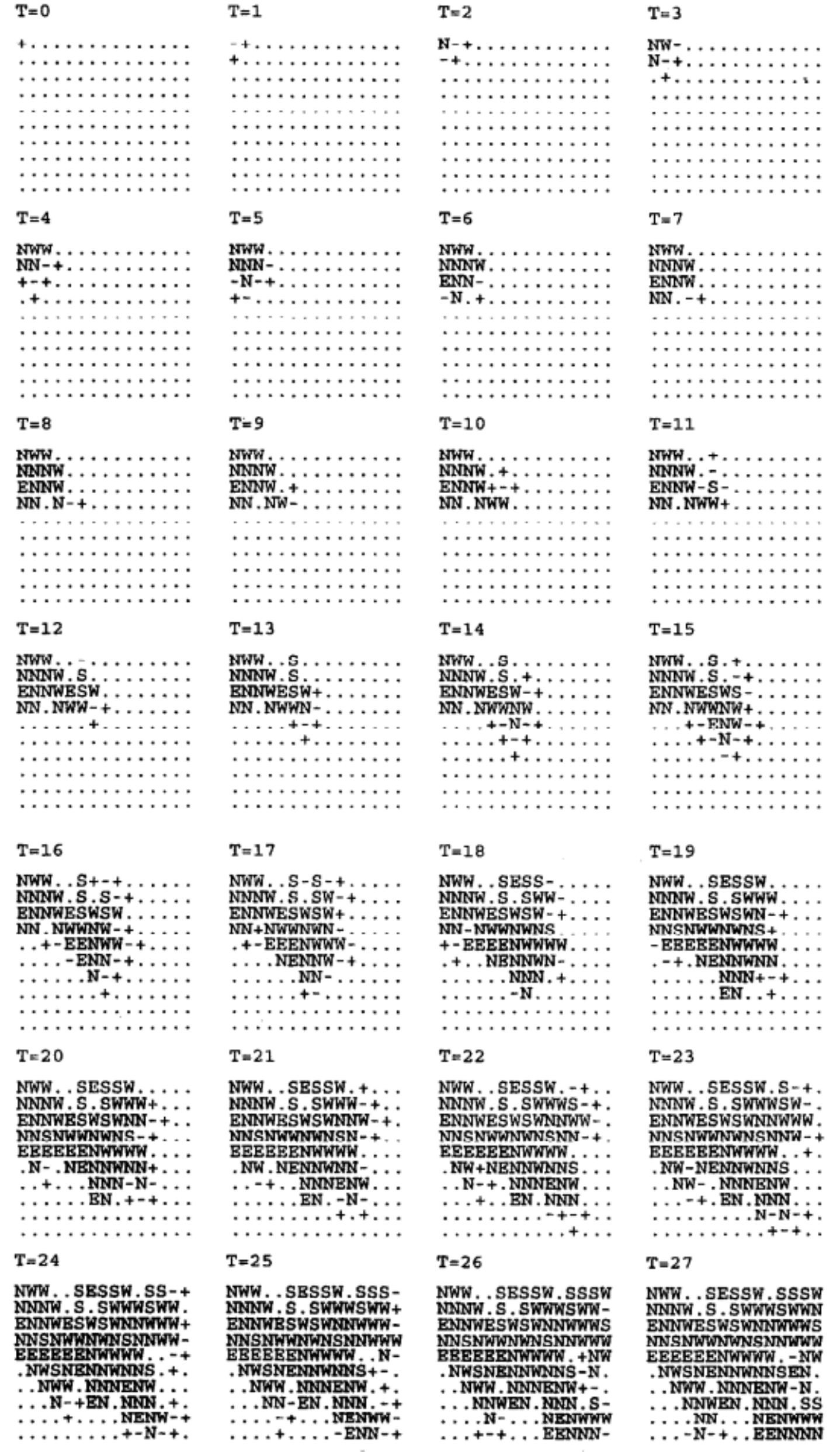}
	\caption{A CA evolution in the computation $S^3DSP$. Symbol $-$ means state $\#$. A state pointer for cell $x$ is shown in the figure only if $x$ was in state $\#$ at least two times (adopted from \cite{adamatzky1996computation}).}
	\label{fig:AndyEvolution}
\end{figure}

\begin{figure}[htbp]
	\centering
	\includegraphics[scale=0.4]{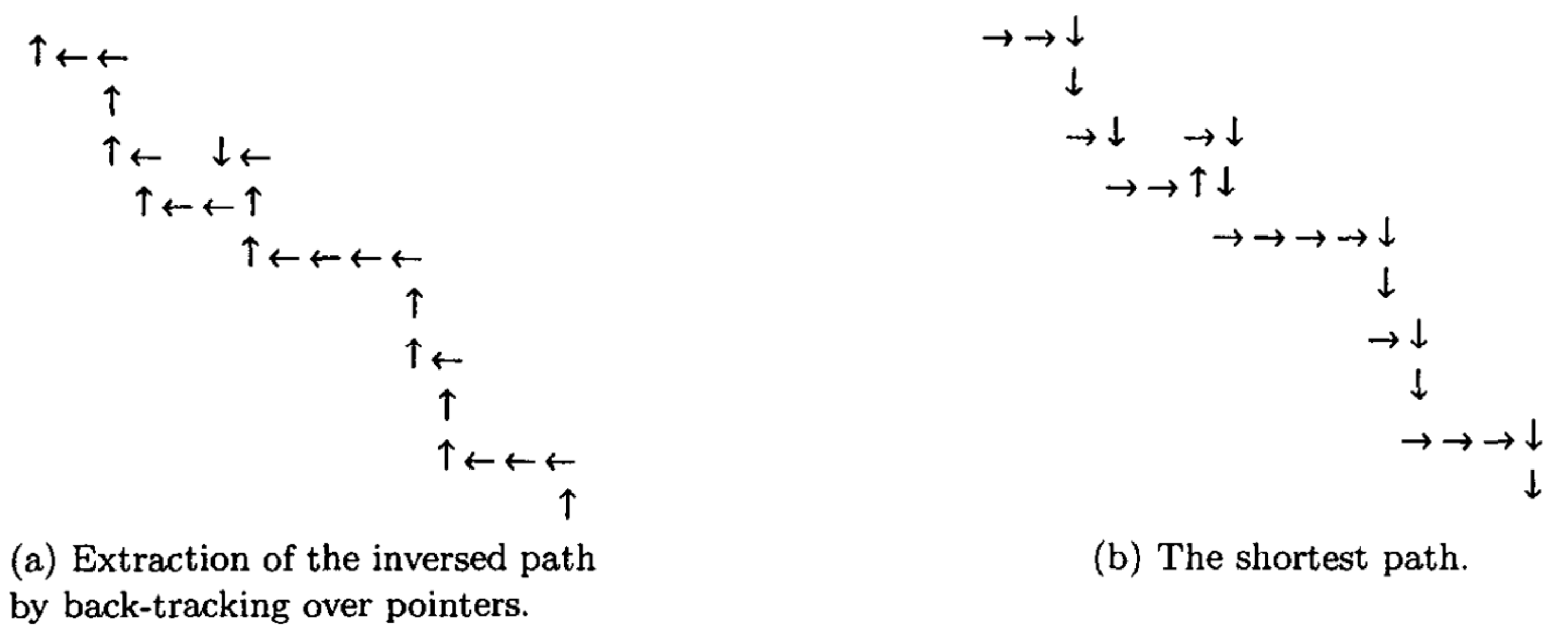}
	\caption{The results of computation of $S^3DSP$ adopted from \cite{adamatzky1996computation}.}
	\label{fig:AndyFirstRes}
\end{figure}

An example can be shown in Fig. \ref{fig:AndyGraph}. The goal is the solution of a $S^3DSP$ in a 2-d grid where the edges can take states $\infty$ or 0 and the source vertex is on the upper left and destination vertex on the bottom right. $Q$ can take values $\{\bullet,+,\#\}$, $Y$ can take values $\{N,W,S,E,\lambda\}$ and $W$ can take values $(w_{xN},w_{xW},w_{xS},w_{xE})$. Symbols $N, W, S, E$ are the indices for the northern, western, southern and eastern neighbors of the cell. The initial conditions are: $x^0_s=+$, $\forall x \in L, x\neq x_s : x^0:=\bullet$ and $p_x:=\lambda$. The dynamic of CA evolution is shown in Fig. \ref{fig:AndyEvolution}. The back-traced (a) and extracted (b) paths can be found in Fig. \ref{fig:AndyFirstRes}.

The longest path in $G$ consists of $n-1$ nodes. So, a CA of $n$ cells, four neighbors and nine states that models a 2-d $G$ graph of $n$ nodes, some cut-off edges and edge weights $\{\infty,0\}$ can compute the shortest path in $O(n)$ computation time and APSP in $O(n^2)$, respectively.

Assuming a 2-d CA of $n$ cells is used, each of which has four neighbors and $6+5n$ states modeling a 2-d rectangular grid where the edges of $G$ graph can be of weight $\{0,...,\nu,\infty \}$, the $S^3DSP$ can be solved in $O(\nu n)$ upper time.

\begin{figure}[htbp]
	\centering
	\includegraphics[scale=0.3]{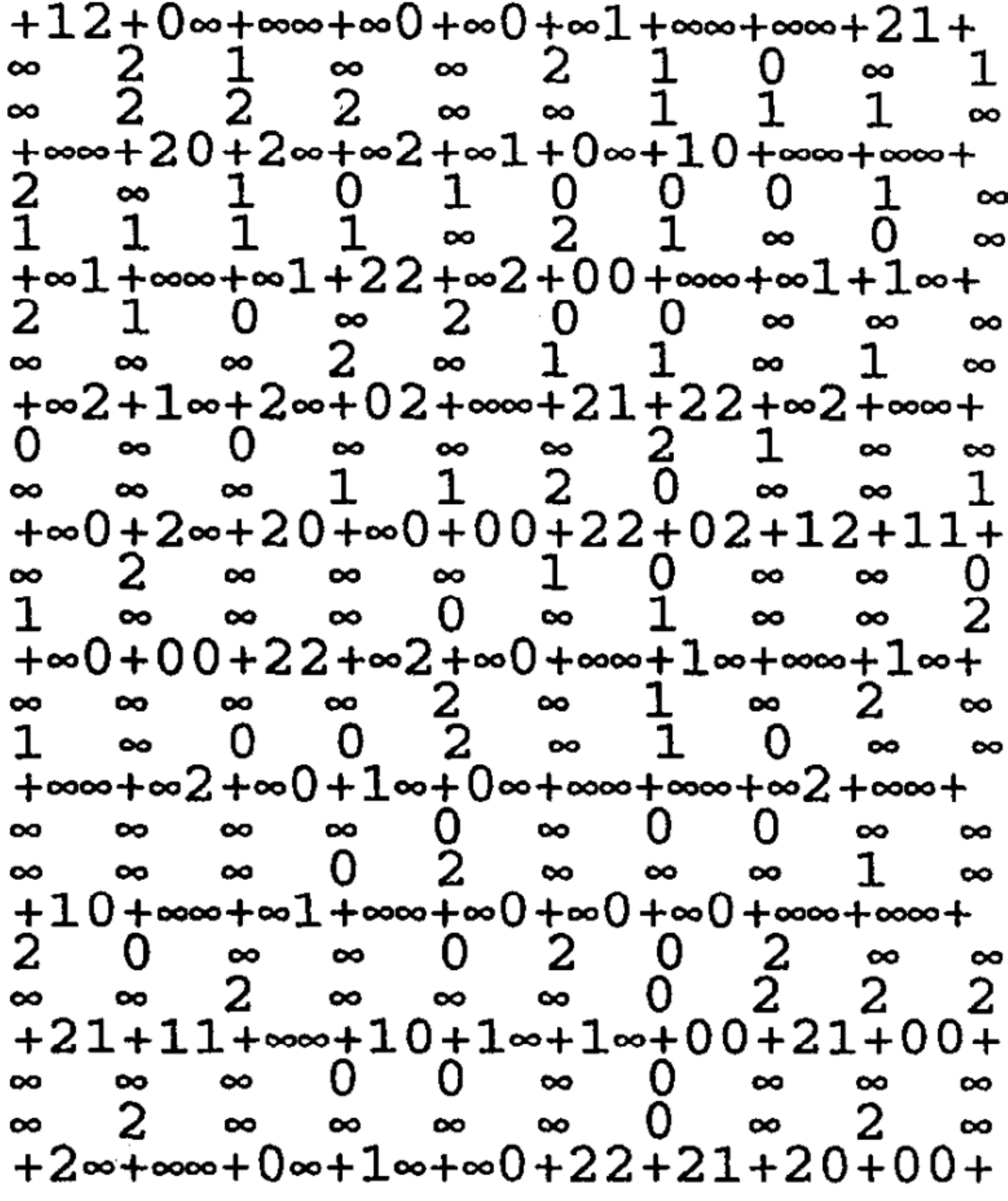}
	\caption{An oriented planar graph $G$ every vertex of which has indegree 4. The edges are weighted with elements of set $\{0,1,2,\infty\}$. The left upper vertex is a source vertex and the right lower is a destination one (adopted from \cite{adamatzky1996computation}).}
	\label{fig:AndySecondMaze}
\end{figure}

\begin{figure}[htbp]
	\centering
	\includegraphics[scale=0.8]{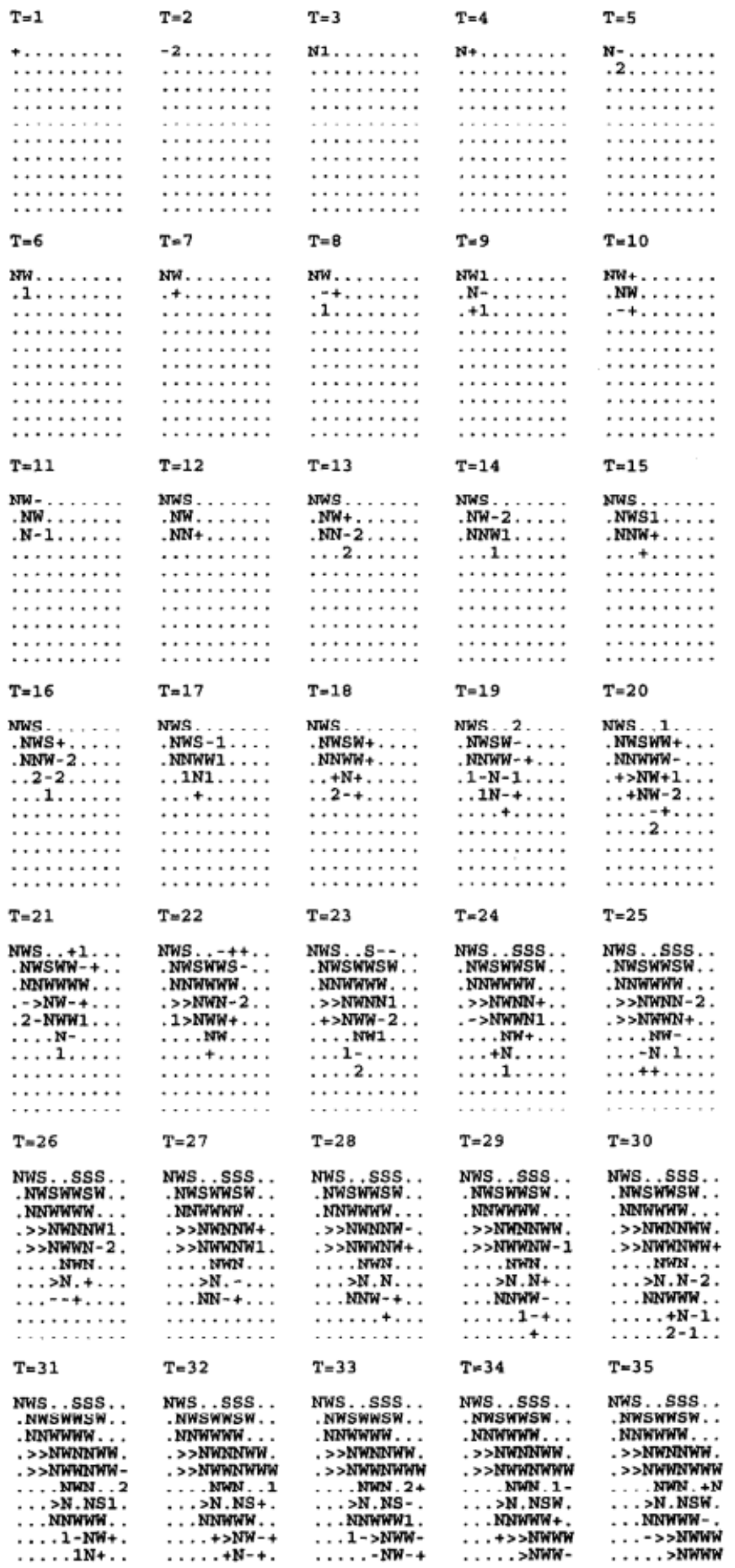}
	\caption{An evolution of CA which computed the $S^3DSP$ on a graph with weighted edges. Symbol $-$ means $\#$. A state of the pointer for cell $x$ is shown in figure only if $x$ was in state $\#$ at least two times (adopted from \cite{adamatzky1996computation}).}
	\label{fig:AndySecondEvolution}
\end{figure}

\begin{figure}[htbp]
	\centering
	\includegraphics[scale=0.2]{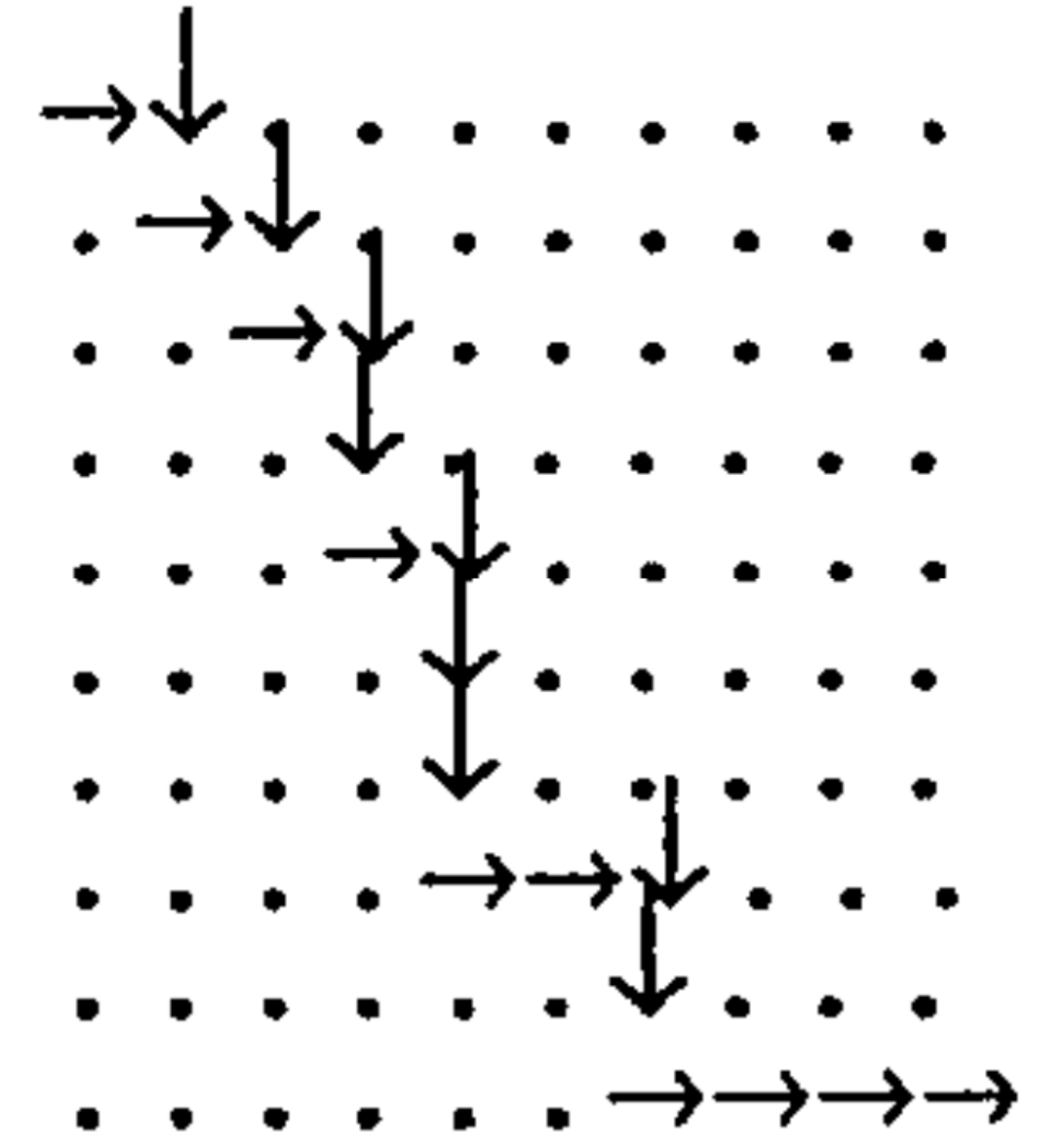}
	\caption{The shortest path (adopted from \cite{adamatzky1996computation}).}
	\label{fig:AndySecondRes}
\end{figure}

The same example is used as in Fig. \ref{fig:AndyGraph}, with set $W$ taking values $\{0,1,2,\infty\}$. The graph with the edges' weight can be seen in Fig. \ref{fig:AndySecondMaze}. If the longest path consists of $n$ vertices and the delay of state transition is $\nu$ for any cell, then $O(\nu n)$ is an upper bound on the time for computation $S^3P$. If a CA uses $k$ neighbors and $O(\nu k)$ states then the $S^3DSP$ can be solved in $O((\nu+k)n)$ upper time. An $S^3P$ can be extracted in $O(n^2)$ upper time. The evolution of the above example can be found in Fig. \ref{fig:AndySecondEvolution} and the result in Fig. \ref{fig:AndySecondRes}.

After the initial introduction of the CA notion in the shortest path problem by Adamatzky \cite{adamatzky1996computation}, an increasing interest of the research community in the specific field was declared. In particular in 2004, more than a decade afterwards, Wu and Xue \cite{WuXue} tried to extend the CA model for shortest path calculation defining properly the cell state and providing suitable cellular evolution to dictate cell states evolution mainly focusing on the appropriate node order as resulted from the proposed CA algorithm. An update of the presented study was published a couple of years later by Li \emph{et al.} \cite{Li2006} with the addition of the cell state turnover and Sun and Dai, based on this CA algorithm proposed the subtraction of the least surplus weight to advance once again the CA based algorithms for shortest path computation \cite{Sun2009} as mentioned in Wang \emph{et al.} paper \cite{Wang2012}. In this last work, Wang \emph{et al.} selected to adjust the cell state set by combining breeding and mature states, while trying to improve the resulting parallelism and at the same time, recording manner of cellular state turnover is modified to record all information sources \cite{Wang2012}.

Moreover, in 2010, Wang \cite{3D} studied the shortest path solution on a three-dimensional surface using CAs. On the 2-d space, a straight line is the shortest distance between two points, but for a complex 3-d surface such as a path over a mountain, the shortest walking path between the starting point and destination can not be a simple straight line. It is a more complex problem to find the shortest path on the complex 3-d surface. Such an approach has a considerable arbitrariness, and it is hard to find the best route. There are also obstacles like hills, rivers, lakes which block the routes from the source to the destination point. 3-d position data of the study area are imported, including the plane coordinates and elevation values of the starting point \textit{A} and end point \textit{B}. The problem is seeking the shortest path between \textit{A} point and \textit{B} point. The distance between \textit{A} and \textit{B} point was divided into $n$ equal portions, and the vertical profile is made over each equal point, so that each vertical profile intersects 3-d surface, $n$ profile curves are derived. Then, each profile curve is equal to $m$ number of points according to horizontal distance, so that the path search problem of 3-d surface is transformed into a discrete optimization problem through gridding.

\section{A Cellular Automata Algorithm Based on Lee's Algorithm}
\label{CALee}

The Lee algorithm \cite{lee} \index{Lee algorithm} is a well known fundamental routing algorithm that can find the shortest path between two points in a grid. The task was to find a CA similar to the Lee algorithm that uses a small number of states, independently of the grid size. Such an algorithm was found in \cite{HUSSAIN} and then further described in \cite{ACRI96}. It was developed when the expressive power
of the cellular description language CDL \cite{PACT95,MPCS96a} \index{CDL} was explored for different applications.

\subsection{Lee Algorithm}
A very well known approach to routing problems is the Lee algorithm \cite{lee}. Its purpose is to find an optimal path between two points on a regular grid. Each point of the grid is associated with a weight. The algorithm finds the path on the grid with the lowest sum of weights. By adjusting the weights of the grid points the user has some control over what is supposed to be  an optimal path. Let us consider an example:
The user simply searches for the shortest path between two points. In this case the user specifies the weight one for all grid points and the algorithm will find the path with the lowest number of grid
points. This is the shortest path between the two chosen points \texttt{S} and \texttt{E}. In another example the user looks for a path that crosses the already existing paths as few as possible. In this case the user assigns a very high weight to all points of the existing paths and a very low weight to all other grid points. The algorithm will then find the path with as few crossings as possible.

The algorithm works in two phases, (1) wave propagation from \texttt{S} to \texttt{E}, and (2) building a path by backtracing from \texttt{E} to \texttt{S}.

%====================================lee1
\begin{figure}[htb]

\begin{em}
for all grid points i do

 \hspace{5mm}  acw(i) := infinity;
 
acw(starting point) :=0

steady := false

while not steady do

\hspace{5mm}  
   steady := true;
	
\hspace{5mm}  
     for all grid points i do
	
\hspace{10mm} 
      MinNeighbour = min(acw(j)) where j in neighbours(i);
			
\hspace{10mm}       
      acw(i) := MinNeighbour + weight(i) of this point;
			
\hspace{10mm}        
      if acw(i) has changed then steady := false;

\end{em}

\caption
{First phase of the original Lee algorithm.
 %At each time step the minimum of the neighbors' acummulated weight acw plus the own weight is added to acw-
}
\label{lee1}
\end{figure}
%====================================lee1

\begin{figure}[htb]
\centering
\includegraphics[width=11cm]{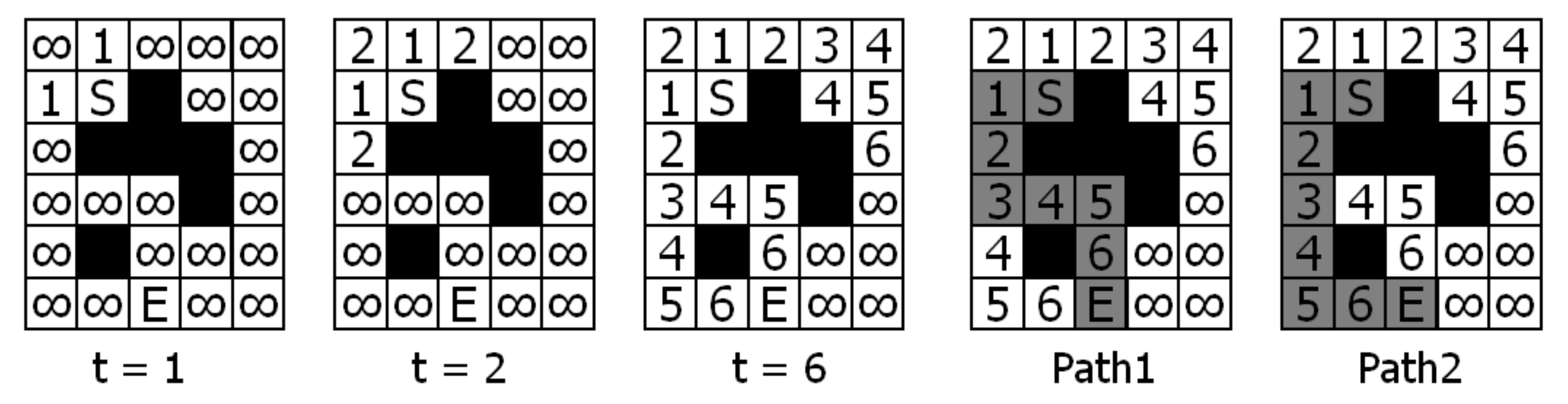}
\caption
{Original Lee algorithm. A wave propagates from \texttt{S} to \texttt{E}. At each time step the cell's weight (one here) is added to the minimum of the neighbours' accumulated weights. When \texttt{E} is reached the path is backtraced along the neighbours with the minimal accumulated weights. Several alternative shortest paths are possible. 
}
\label{lee-original}
\end{figure}

In the first phase (Fig. \ref{lee1}) the accumulated weights (acw) for each node relative to the starting point are computed. All free cells are initialized to infinity. The accumulated weight for the starting point \texttt{S} is initialized to 0. Fig. \ref{lee-original} shows for a sample grid the calculation of the accumulated weights. The weight of all grid points is one in this case. The wave reaches the end point at time step $t=6$.

In the second phase the actual path is established. For this purpose the algorithm 'walks' back from the end point \texttt{E} to the starting point \texttt{S}. At each step the neighbour with the smallest accumulated weight is selected as part of the path.  

% towards a neighbour which has. This step is repeated until the algorithm reaches the starting point. 

Note, that there are several possibilities to build a path from the end point to the start point. At the end point you can either go up or left (Path1 and Path2 in Fig. \ref{lee-original}, respectively) for this example. 

%====================================lee-original
%====================================lee-original

Obstacles can be modeled by infinite weights at grid points. Since then the sum of the local weight and the accumulated weight of a neighbour will always be infinite, such grid points may never become part of the
path.

At a first glance this algorithm looks like it perfectly fits onto a CA. Unfortunately the number of states required to perform the algorithm is related to the longest path or more precisely to
the largest accumulated weight that can occur. Thus we decided to develop a version of the algorithm\cite{HUSSAIN,ACRI96} which has a constant number of states. This version can only handle the shortest path problem with a unified weight at all grid points.

%%%%%%%%%%%%%%%%%%%%%%%%%%%%%%%%%%%%%%%%%%%%%%%%%%%%%%%%%%%%%%%%%%%%%
\subsection{CA based Lee Algorithm}

The accumulated weights in the Lee algorithm are needed to find the shortest path. Instead of storing the accumulated weights % we could also
we stored the direction in which we have to move in order to return back to the starting point. With these wave marks instead of the accumulated weights the algorithm requires only a constant set of states. Of course, we can not handle problems with arbitrary weights at the grid points.

We present the modified algorithm in \index{CDL} CDL (Cellular Description Language). \index{cellular description language} This language has been developed to describe CAs in a concise and readable way, independent of the target architecture. Compilers were built that could translate CDL into C functions for almost any software architecture and into Boolean equations for hardware synthesis, like the CEPRA family \cite{CERBAL}.

%====================================lee-algo
\begin{figure}[htb]
\centering
\includegraphics[width=11.5cm]{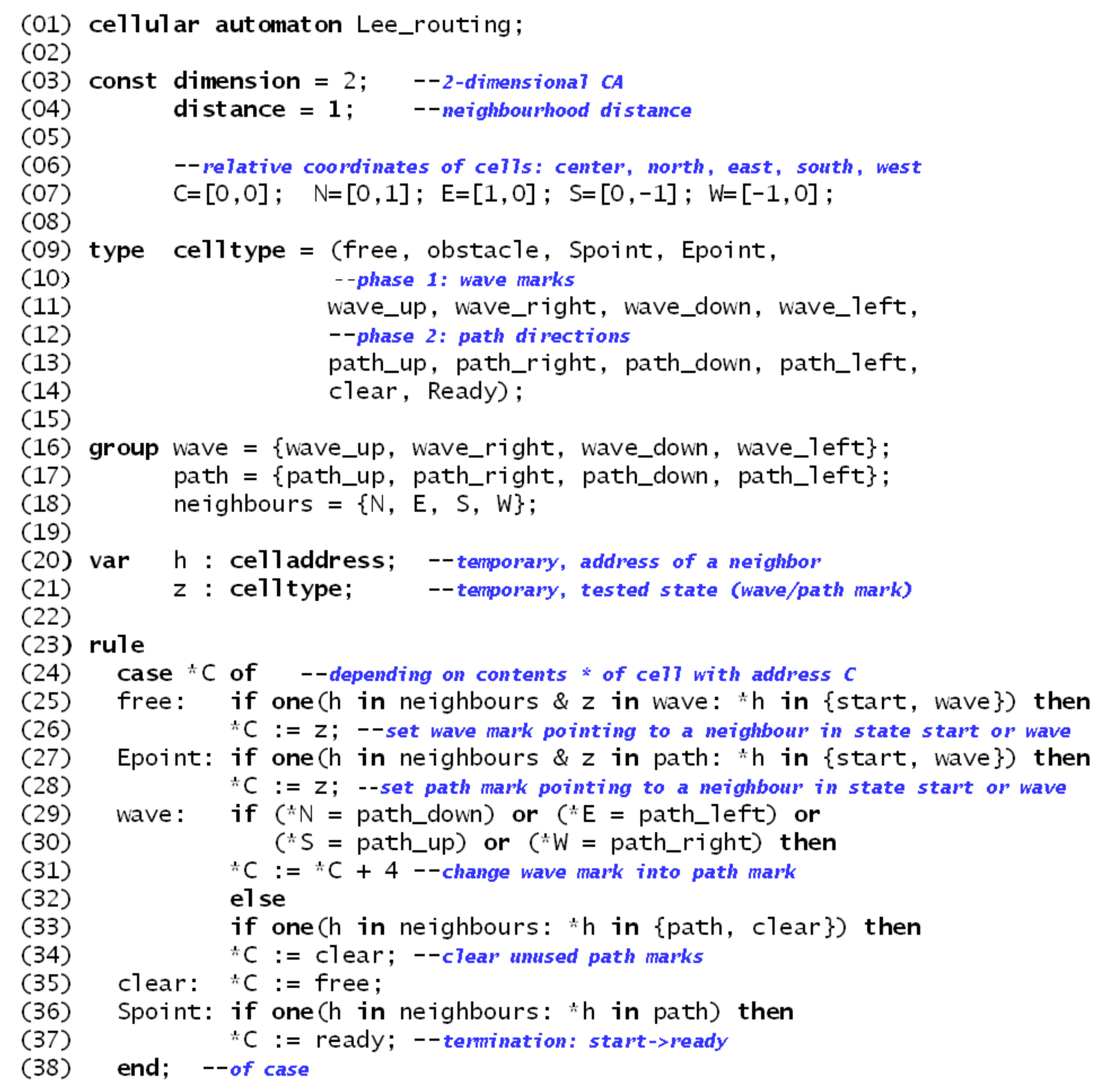}
\caption
{The shortest path algorithm described in the language CDL.
}
\label{lee-algo}
\end{figure}
%====================================lee-algo

\begin{figure}[htb]
\centering
\includegraphics[width=11.5cm]{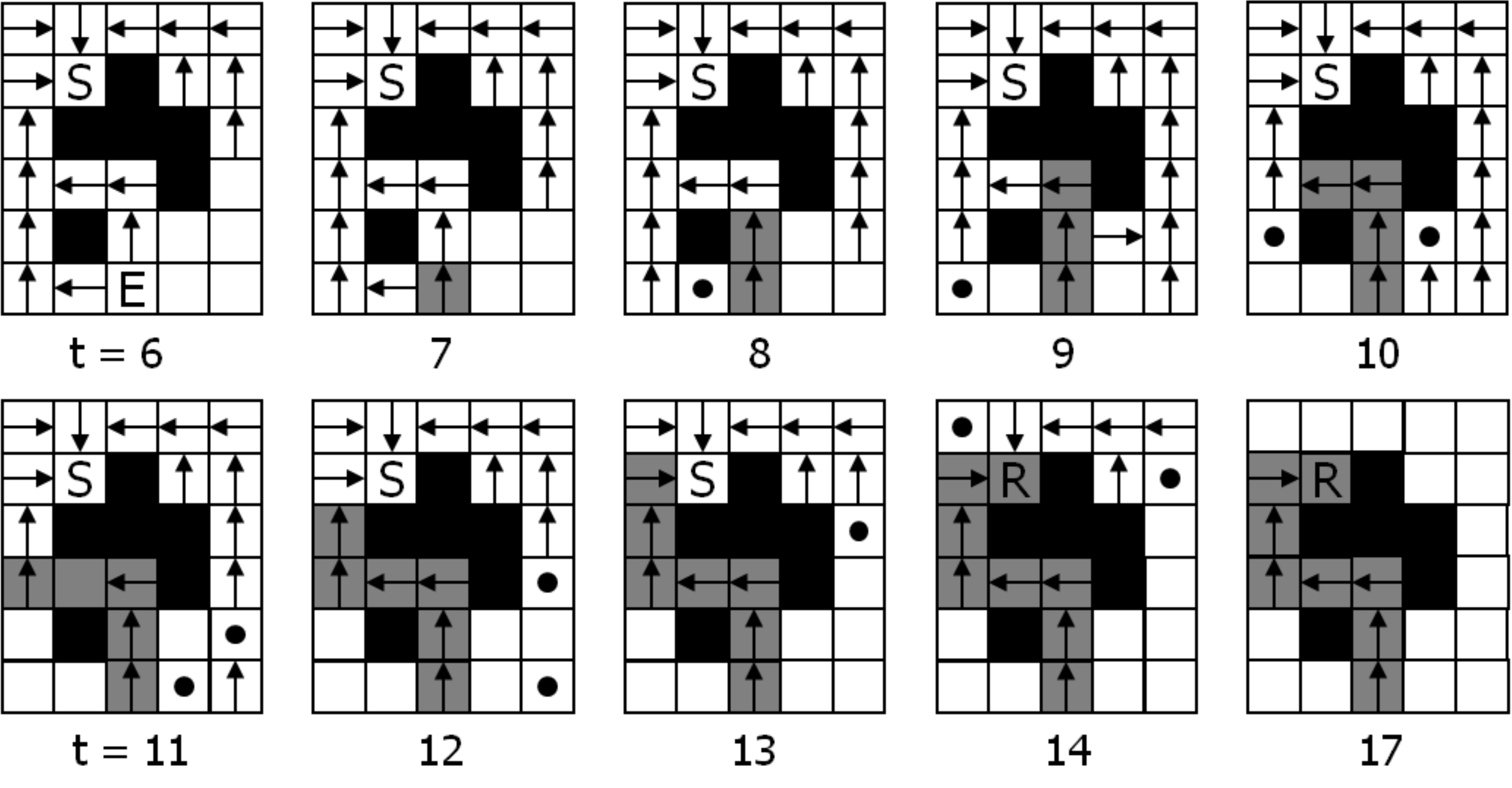}
\caption
{A simulation of the algorithm. 
$(t=6)$ Wave marks at the end of phase one. 
$(t=8)$ A first unused wave mark changes into the state \texttt{clear} ($\bullet$). 
$(t=13)$ Path is constructed. 
$(t=14)$ Starting point changes into the \texttt{ready} state (R). 
$(t=17)$ All unused wave marks are deleted. 
}
\label{lee-simulation}
\end{figure}

At the beginning of the first phase all cells are in the \texttt{free} state, except for the starting and end point. In the first phase, all cells check whether there is any cell in the neighbourhood that already
has a wave mark. If a wave mark is found, the cell itself becomes a wave mark towards the already marked cell. This is performed in lines 25 and 26 of Fig. \ref{lee-algo}. The \texttt{one} function successively assigns all the elements of the groups \texttt{neighbours} and \texttt{wave} to the temporary variables \texttt{h} (a cell address) and \texttt{z} (a state). For each assignment the condition following the colon is checked. The evaluation of the \texttt{one} function stops if an assignment is found, that satisfies the condition, i.e. the corresponding neighbour is in state \texttt{start} or in any of the states \texttt{wave}. The first assignment is \verb+h=N+ and \verb+z=wave_up+. The assignment to \texttt{z} is only used for its side effect to store temporarily the wave state corresponding to the neighbour being investigated. If the east neighbour is currently investigated this cell state will change to \verb+wave_right+ since it must point to this neighbour. Figure \ref{lee-simulation} at $t=6$ shows a sample grid at the end of phase one. The wave marks are symbolized by small arrows. The black squares are obstacles. We had to introduce a special state to model obstacles, since we do not have weights at the grid points.

%====================================lee-simulation
%====================================lee-simulation

Phase one ends when the end point is reached. Now the path is built backward towards the start point along the wave marks (lines 29--34). If a cell is one of the \texttt{wave} states and it sees a neighbour cell that is a path towards this cell, then this cell becomes a path in the direction of its previous mark (Fig. \ref{lee-simulation}, $t=7 - 13$). This is done in the CDL program by adding four to the enumeration element; e.g. \verb+wave_down+ becomes \verb+path_down+. When the starting point is reached, its state changes from \texttt{S} to \texttt{R}. The ready state \texttt{R} signalizes the termination of phase two. 

All unnecessary wave marks have to be cleared in order to allow subsequent routing passes. For this purpose all cells that see a neighbour cell which is a path not pointing towards this cell are cleared. Such a cell will never become part of the path. Also all cells are cleared that see a neighbour in the clear state. Since the building of the path moves along the shortest path, it is impossible that a cell in clear state could reach a cell which will be in the path but is not yet part of it. A cell in state \texttt{clear} will change to state \texttt{free} in the following time step.

The time complexity for the first and second phase is \textit{O(p)}, where $p$ is the path length. For a $n \times n$ square grid, the maximal path length is $2n-1$ if there are no obstacles, and $O(n^2)$ if there are obstacles (spiral like path). The algorithm requires only 14 states and thus can be very easily realized in hardware.

%%%\textbf{Summary.}
Based on the Lee algorithm, a CA algorithm with 14 states was designed that is able to route a shortest path between to nodes of a grid. The algorithm is independent of the grid size and needs only 14 states per cell. The time complexity is \textit{O(p)}, where $p$ is the path length.

\section{Learning Automata}
\label{sec:DLA}

There are shortest path problems in which the lengths of the edges in a graph are allowed to be random. This makes the problem more difficult. A stochastic graph $G$ is defined by a triple $G =\langle V, E, F \rangle$, where $V$ represents the set of nodes, $E$ specifies a set of edges and matrix $F$ is the probability distribution describing the statistics of edge lengths. In stochastic graph $G$, a path $\pi_i$ with length of $n_i$ nodes and expected length of $L_{\pi_i}$ from source node $V_{source}$ to destination node $V_{dest}$ is defined as an ordering \{$\pi_{i,1}, \pi_{i,2}... \pi_{i,n_i}$\} of nodes. $V_{\pi_{i,1}} = V_{source}$ and $V_{\pi_{i,n_i}}=V_{dest}$ are source and destination nodes, respectively and all the intermediates are nodes in path $\pi_i$. Assume that there are $r$ distinct paths between $V_{source}$ and $V_{dest}$. The shortest path between source node and destination node is defined as a path with minimum expected length. Stochastic shortest path problems can be grouped in two main classes: the first class aims to find a priori solution that minimizes the expected lengths, while the second one computes an online solution that allows decisions to be made at various stages.

In 2006, Beigy and Meybodi \cite{Beigy} introduced a network of Learning Automata (LAs), \index{learning automata} which were called Distributed Learning Automata (DLAs). \index{distributed learning automata} More specifically, an automaton acting in an unknown random environment and improving its performance in some specified manner, is referred to as a learning automaton (LA). DLAs is a network of automata which collectively cooperate to solve a particular problem. A DLA can be modeled by a directed graph in which the set of nodes of a graph constitute the set of automata and the set of outgoing edges for each node constitutes the set of actions for corresponding automaton. When an automaton selects one of its actions, another automaton on the other end of edge, corresponding to the selected action, will be activated. They used this tool to propose some iterative algorithms and solve stochastic shortest path problem. In these algorithms, at each stage DLAs determine which edges to be sampled. This sampling method may result in decreasing unnecessary samples and hence decreasing the running time of algorithms. The automata approach to learning involves the determination of an optimal action from a set of allowable actions. Automaton selects an action from its finite set of actions, which serves as the input to the environment, which in turn emits a stochastic response at a specific time. The environment penalizes or rewards an action of the automaton with a penalty/reward probability. The state of the automaton is updated and a new action is chosen at the next time step. 

In their algorithm a network of learning automata which is isomorphic to the input graph is created. Each node in the network represents a LA and the actions of this node in the LA is the outgoing edge of this node. Then, at the stage $k$, source automaton, which represents the source node in the graph, chooses one of its actions based on its action probability vector. If the action is $a_m$ the automaton $A_m$ is also activated on the other end of edge $(s, m)$. The process of choosing an action and activating an automaton is repeated until the destination automaton is reached or for some reason moving along the edges of the graph is not possible or the number of visited nodes exceeds the number of nodes in the graph. After destination automaton is reached, the length of the traversed path is computed and then compared with a quantity called dynamic threshold. 

Depending on the result of the comparison all the LAs (except the destination learning automaton) along the traversed path update their action probabilities. Updating is done in direction from source to destination or vice verse. If length of the traversed path is less than or equal to the dynamic threshold then all LAs along that path receive reward and if length of the traversed path is greater than the dynamic threshold or the destination node is not reached, then activated automata receive penalty. The process of traveling from the source LA to the destination LA is repeated until the stopping condition is reached which at this point the last traversed path is the path which has the minimum expected length among all paths from the source to the destination.

\section{Shortest path based on Cellular Automata Algorithm for Computer Networks}
\label{sec:net}

As already mentioned in the Introductory Chapter the problem of finding the shortest path (SP) from a single source to a single destination in a graph arises as a sub-problem to many broader problems. In general, different path metrics are used for different application. For example, in communication systems, if each link cost is 1, then the minimum number of hubs is found. However, cost can also represent the propagation delay, the link congestion or the reliability of each link. In the latter case, if the individual communication links operate independently, then the problem can be stated as to find what path has the maximum reliability.

Here we focus on the computer networks and we present how the aforementioned problem is confronted and solved by the CAs approach. More specifically, Mardiris \emph{et al.} \cite{Mardiris} presented an interactive tool that offers automated modeling with the assistance of a dynamic and user friendly graphical environment, called Net\_CA for modeling and simulation of computer networks based on CAs. More specifically, a 2-d NaSch \cite{Nagel} CA computer network model was developed and several computer networks were simulated, while algorithms for connectivity evaluation, system reliability evaluation and shortest path computation in a computer network have also been implemented. The proposed system also produced automatically synthesizable VHDL code leading to the parallel hardware implementation of the above CA algorithms rendering Net\_CA as a very reliable and fast simulator for wireless networks, ad hoc networks and, generally, for low connection reliability networks.

In regards to the shortest path algorithm as expressed in CAs and applied for computer networks, let $G=(N, A)$ be a network, where $N$ is the set of $n$ nodes, $A\subseteq{N\times{N}}$ is the set of connections and $L_i$ is the neighborhood of the node $i$. That is, each node $i$ is mapped to a cell whose neighborhood is the set of nodes connected to it by its input connections. Associated with each connection $(p,q)\in{A}$ is a non-negative number; $C_{pq}$ stands for the cost of connection from node $p$ to node $q$. Non-existing connection costs are set to infinity. Let $P_{st}$ be a path from a source node $s$ to a destination node $t$, defined as the set of consecutive connected nodes: $P_{st}={\left\{s; n_1; n_2; \ldots; n_i; t\right\}}$. The state of each node, at each time step $t_s$, is represented by a vector with two entries $V_i(t_s)={\left\{V^1_i(t_s),V^2_i(t_s)\right\}}$: the first component is a pointer to the previous node in the path, while the second is the cost of the partial path up to node $i$. $V^1_i(t_s)$ is not necessary for evaluating the shortest path length, but it is used only for indicating the shortest path itself.

The evolution of CA algorithm is given by the following equation:

\begin{equation}
\label{eq:mardiris}
V^1_i(t_s)=\left\{k,\left(V^2_i(t_s)+C_{k,i}\right)\right\}
\end{equation}

\begin{figure}[h!]
	\centering
	\includegraphics{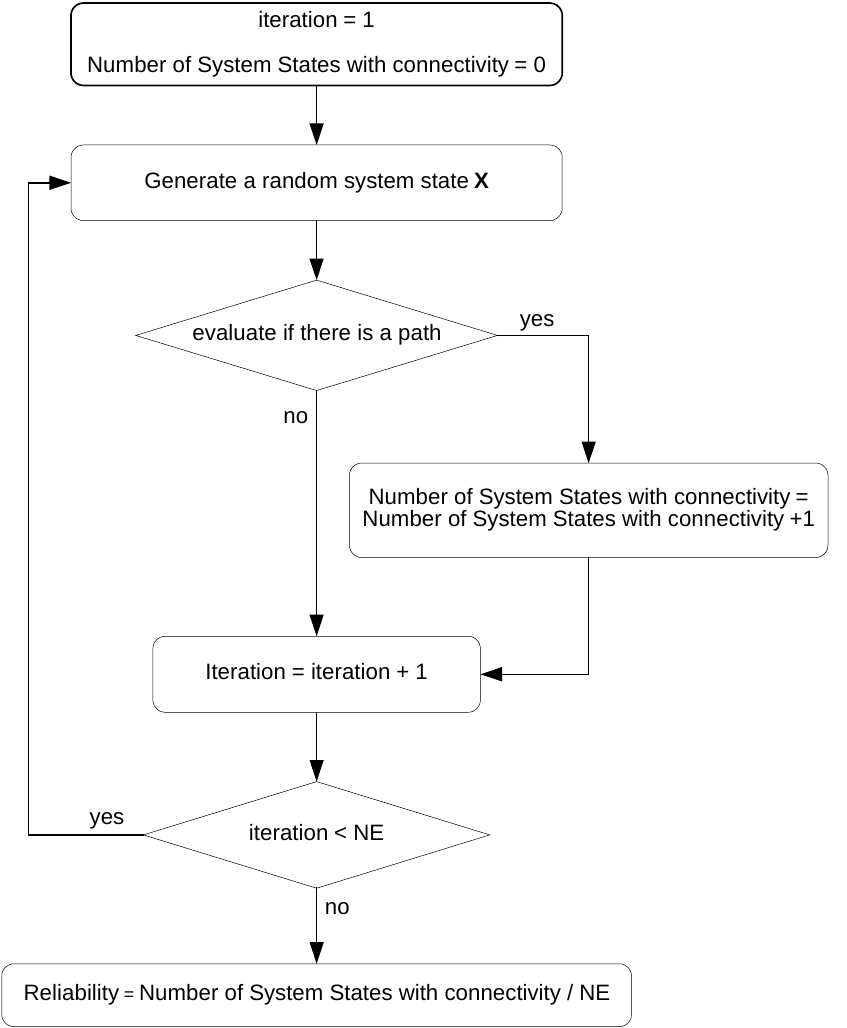}
	\caption{The CA algorithm flowchart of shortest path computations (adopted from \cite{Mardiris}).}
	\label{fig:flowchart}
\end{figure}

The flowchart of the described CA algorithm is shown in Fig. \ref{fig:flowchart}. The minimum $s-t$ path cost is the second component of the state vector of node $t$. In case this value is infinity, then there exists no $s-t$ path.

\begin{figure}[h!]
	\centering
	\includegraphics{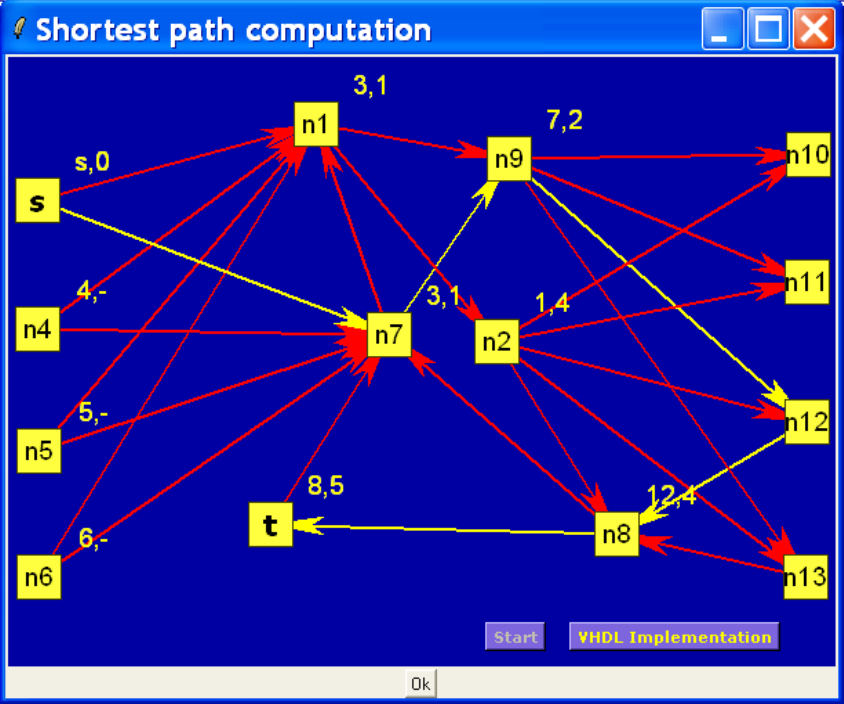}
	\caption{The final simulation screen of the Net\_CA system during the execution of the shortest path algorithm between nodes $s$ and $t$ (adopted from \cite{Mardiris}).}
	\label{fig:NET}
\end{figure}

The results of the implementation of the presented shortest path computation algorithm to the proposed Net\_CA system are depicted in Fig. \ref{fig:NET}. As before the user defines the starting topology. During the execution of the aforementioned algorithm each node is described by a pair of values, one for its number (name) according to the cost of the connection up to it, and the other for the minimum cost of connection of the starting node to the examined node. After the execution of the algorithm the shortest connectivity path between nodes $s$ and $t$, if it exists, is colored yellow.

\section{Shortest Path Definition based on Cellular Structures for Coordinated Motion in Swarm Robotics}
\label{sec:swarmRobots}

In robotics, the applied shortest path solvers must consider various constraints regarding both the environment and the utilized robot configurations. Robot navigation involves the determination of a continuous motion for a robot towards a goal location. Depending on the available information, the path planners must consider the presence of obstacles in order to avoid potential collisions. A priori knowledge of the configuration area status involves static obstacles which results to simple solutions of the path-planning problem. A global shortest path is extracted for the robot, which involves also collision avoidance scenarios. The complexity of such solutions increases in cases where no information is available in advance (dynamic environments) and therefore, they should extract the robot's motion in real-time. Robot's motions are recalculated at every time step in order to both avoid the detected objects and follow the defined shortest path. The periodic motions of a robot could also be affected by processes of coordination in multirobot systems. The required collective behavior that emerges from the interactions between the robots significantly impress the results of the path planner and its overall complexity. Cooperative robotic teams are extensively utilized for accomplishing additional tasks such as exploration \cite{nieto2014coordination}, search and rescue \cite{macwan2015multirobot} and formation control \cite{nascimento2016multi}.

Several methods have been proposed for solving the shortest path problem in robot navigation, both for a single or multiple robots and for static or dynamic environments. Visibility graphs have been exploited to identify Euclidean shortest paths among a set of polygonal obstacles in the plane \cite{moghaddam2016planning}. The method applies a recursive process capable of solving dynamic navigation problems for a single robot. Collective behavior will dramatically increase the complexity of the approach due to recursiveness rendering the method improper for computing shortest paths in multirobot systems. Moreover, a modified version of potential fields has been proposed in order to consider both static and dynamic obstacles \cite{montiel2015path}. The method combines an Artificial Potential Field technique with a Bacterial Evolutionary Algorithm to reduce the extraction time of the optimal path. Despite the fast computation and the accurate results, the resulted time remains in high levels due to its dependency with the utilized robot configurations making its implementation on swarm robots unfeasible. Aiming at smoother transitions, heuristic based algorithms were introduced as potential path planning solvers. For example, an extended version of a D-star algorithm \cite{stentz1995focussed} was proposed in \cite{ferguson2006using}. The method applies a bilinear interpolation function to compute the required motion fragments resulted by the vertex expansion. In order to reduce time complexity and define smoother transitions, a mathematical model inspired by Physarum polycephalum along with a heuristic rule function were proposed to solve the shortest path problem \cite{zhang2014rapid}. The method extracts accurate results in limited time amounts nonetheless; it could not be implemented in low resources systems.

On the contrary, cell decomposition techniques display low time and computational complexity levels rendering the approaches proper for implementation in swarm robotic systems. The configuration area is partitioned into a lattice grid and every area cell is processed accordingly. For example, free space is retracted onto a Voronoi diagram while the evolution of a CA constructs its structure \cite{tzionas1997collision}. Moreover, a variant of the A-star algorithm, namely Theta-star, was extended in \cite{nash2013any} where the acquired information is propagated along grid edges without constraining the paths to grid edges. The method handles accurately static objects however; the extracted paths for unknown environments are based on assumptions. Moreover, numerous artificial intelligence algorithms were proposed as potential solvers of the shortest path problem in robotics. A fuzzy logic controller was proposed in \cite{pandey2014path} where obstacles of various shapes can be avoided and a single robot can follow the computed shortest path towards its final destination. A fuzzy-based cost function was also exploited by an ant colony optimization for the evaluation process of the potential solutions \cite{garcia2009path}. In addition, various types of artificial neural networks were utilized to extract optimum paths. A Guided Autowave Pulse Coupled Neural Network \cite{syed2014guided} and a Deep Convolutional Neural Network \cite{hwu2017self} were applied to create collision free trajectories for mobile robots. Despite their efficiency, special hardware resources and/or centralized control are required for their implementation to real robotic systems.
On the contrary, cell decomposition techniques display low time and computational complexity levels rendering the approaches proper for implementation in swarm robotic systems. The configuration area is partitioned into a lattice grid and every area cell is processed accordingly. For example, free space is retracted onto a Voronoi diagram while the evolution of a CA constructs its structure \cite{tzionas1997collision}. Moreover, a variant of the A-star algorithm, namely Theta-star, was extended in \cite{nash2013any} where the acquired information is propagated along grid edges without constraining the paths to grid edges. The method handles accurately static objects however; the extracted paths for unknown environments are based on assumptions. Moreover, numerous artificial intelligence algorithms were proposed as potential solvers of the shortest path problem in robotics. A fuzzy logic controller was proposed in \cite{pandey2014path} where obstacles of various shapes can be avoided and a single robot can follow the computed shortest path towards its final destination. A fuzzy-based cost function was also exploited by an ant colony optimization for the evaluation process of the potential solutions \cite{garcia2009path}. In addition, various types of artificial neural networks were utilized to extract optimum paths. A Guided Autowave Pulse Coupled Neural Network \cite{syed2014guided} and a Deep Convolutional Neural Network \cite{hwu2017self} were applied to create collision free trajectories for mobile robots. Despite their efficiency, special hardware resources and/or centralized control are required for their implementation to real robotic systems.

All the aforementioned methods display multiple limitations or specific drawbacks despite their efficiency in defining the required shortest paths for robot navigation. Their vast majority cannot include collective behavior since their complexity and their resource requirements can be increased significantly, even with proper implementation modifications. Several methods have been proposed to overcome such limitations and are specialized in computing the required paths in multiple robot teams. In general, methods that consider collaborations between the robot members present deviations regarding their complexity, which is related to the size of the robotic team and the collaborative tasks. The coordinated movement of a robotic team comprises one of the most widely studied research fields in swarm robotics. For example, a feedback law using Lyapunov type analysis was derived in \cite{mastellone2007remote} for a single robot thus, collision avoidance and tracking of the shortest path are accomplished. The method extends this result to the case of multiple nonholonomic robots. A coordinated control scheme based on leader-follower approach was also proposed to achieve the required formation maneuvers during the robots' transit following their shortest routes \cite{defoort2008sliding}. First and second order sliding mode controllers were used for asymptotically stabilizing the vehicles to a time varying desired formation considering the optimum pathway. In addition, an improved rapidly exploring random tree (RRT) method was proposed in \cite{liu2014dynamic}. The modified RRT considers the kinematics of each mobile robot to extract the corresponding pathways while a dynamic priority strategy was introduced to avoid mutual collisions and retain the formation of the team. Except these approaches, various artificial intelligence based methods were also introduced in mutltirobot systems. A unified framework of a co-evolutionary mechanism and an improved genetic algorithm (GA) were introduced to compute the multiple paths of the team \cite{qu2013improved}. The improved GA converges to the optimum collision free paths while the co-evolution mechanism takes into full account the cooperation between the populations to avoid collisions between mobile robots. Finally, multiple shortest paths can also be defined with the use of artificial bee colony algorithms \cite{liang2015efficient}. 

In general, most of the above methods can produce accurate results for multiple robots while retaining a formation nonetheless; their implementation in real systems is restricted. Most of these algorithms can only operate in simulation environments due to their resource requirements. Here a CA-based path planner is presented for robot teams, which also involves collective behaviors between the robot members in order to define the shortest routes for retaining their formation \cite{IoannidisR,IoannidisApp,IoannidisJCA}. As already briefly commented CAs comprise a simple, yet efficient, computational tool that can be implemented in real systems of low cost miniature robots. CAs were successfully exploited as potential solution of the shortest path problem in \cite{adamatzky1996computation} found earlier in this Chapter and could be denoted as a cell decomposition approach and proper for a single robot application. Marchese has also introduced the use of Spatiotemporal Cellular Automata (SCA) to define the desired shortest path \cite{marchese2015multi}. Three level of maps are introduced where the first two maps reduce the problem of extremely large cell numbers. Limiting the search space to smaller areas and considering the interaction between the robots, motion planning is performed using the SCA. A simpler approach was introduced in \cite{charalampous2014real} where the A-star algorithm was combined with CAs and tested successfully in real world planar environments. More specifically, the finite properties of the A-star algorithm were amalgamated with the CA rules to build up a substantial search strategy \cite{charalampous2012efficient}. The corresponding algorithm's main attribute is that it expands the map state space with respect to time using adaptive time intervals to predict the potential expansion of obstacles.

In the following, a CA-based algorithm is introduced for dynamically extracting the required collision free pathways for every member of a robot team. The presented planner considers also the collective behavior that the robots must display in order to retain their formation. A swarm robotic team must cover a specific space in the configuration area while simultaneously each member must be able to detect and bypass every dynamic obstacle. For cases where a scatter formation is produced due to the existence of an obstacle, the team must be able to recover its initial formation following optimum paths via collaborations. The CA-based algorithm can extract the optimum pathways of every robot towards its final destination point while shortest paths are also computed for recovering the desired formation. In contrast with similar CA-based architectures, the proposed method does not require any type of central control making the system fully autonomous. In addition, the method is applicable to real systems comprised by miniature robots with low resource specifications since the next transit of ever robot depends on only its current location and the states of its adjacent robots. This flexibility and the method's efficiency were tested using different types of formations.

\subsection{Proposed method}
\label{subsec:methodRobots}

As a cell decomposition approach, the configuration area where the teams operates is initially divided into a simple rectangular lattice of identical square cells. Both dimensions of the required lattice are expressed in cells and thus, they depend on the applied cell length. The latter is strictly related to the specifications of the distance sensor that is utilized from a robot in a real system. For the presented model, the desired covered distance in terms of cell numbers comprises the variable that determines the lattice size. Following the CA description, %in \ref{subsub:CATheoritical}, 
variable $D$ is defined based on these requirements. Let $z$ be the cell length and $x \times y$ cells the dimensions of the CA.

Since the dimensions are defined, the set of states should also be defined, meaning variable $Q$. According to the CAs definition, every cell can be denoted with only one discrete state at every evolution step based on the delineated set of states $Q=\left\{0,1,\ldots,q\right\}$. The proposed model includes the use of multiple robots and so, the number of the possible states is relative to the number of the robot cells. Taking this notion one-step further, this robot state is exploited as an identifier for the collective behavior of the team. Assuming that the team includes $r$ robot cells, the final set of states is comprised by three discrete subsets: $CF$ denoting the absence of both obstacle and robot cells (free cells), $CR$ denoting a robot cell and $CO$ the presence of an obstacle. More specifically, every subset of state can be defined as $CF={0}$, $CR={1,2,\ldots,r}$ and $CO={r+1}$. Essentially, in order to avoid the overlapping of the cell states, the equation \ref{eq:EQ4} must be valid:

\begin{equation}
\label{eq:EQ4}
C_F\cap C_R\cap C_O=\{\emptyset\} \& C_F\cup C_R\cup C_O=Q 
\end{equation}

\begin{figure}[htbp]
	\centering
	\includegraphics[scale=0.5]{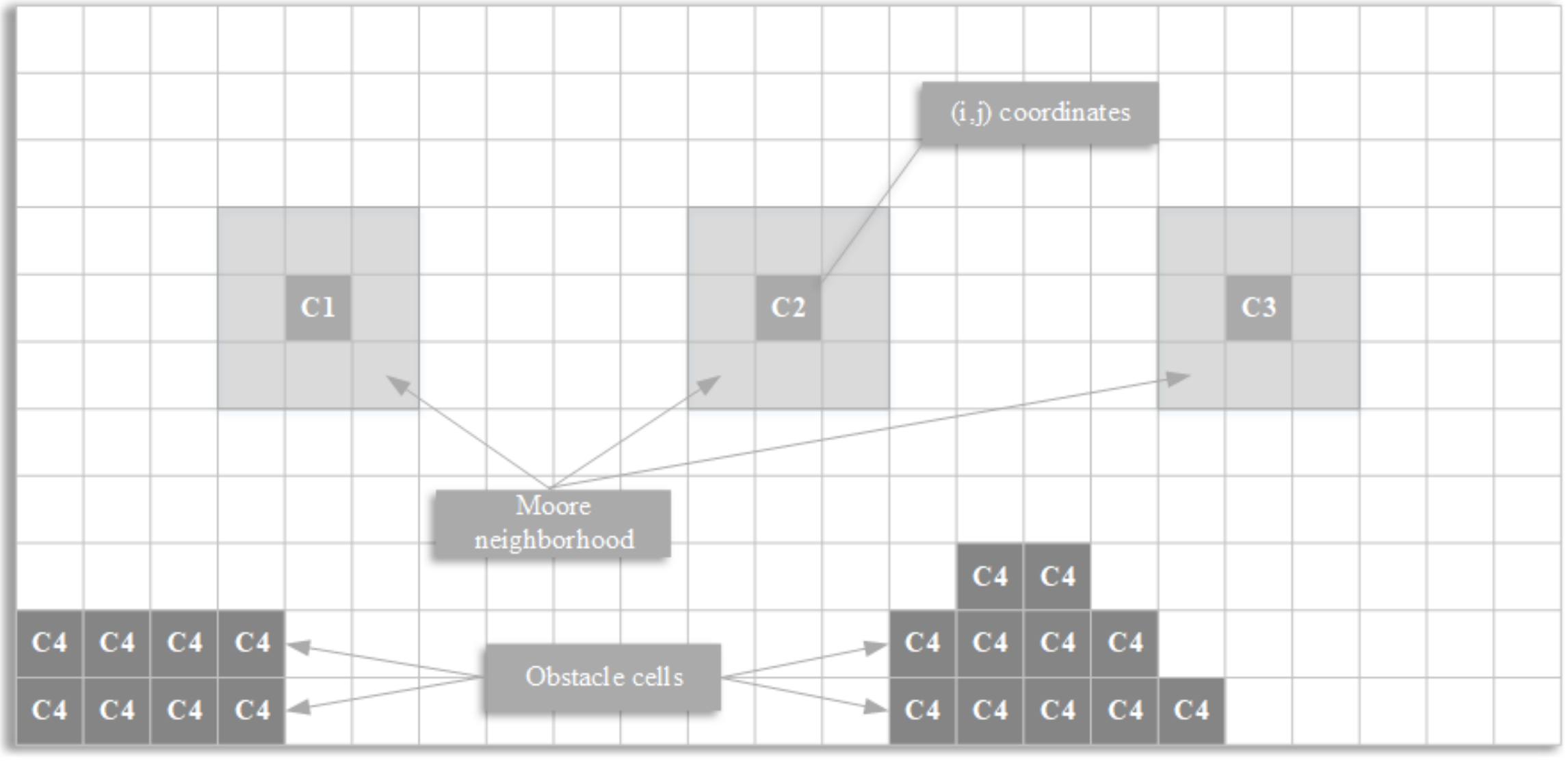}
	\caption{Example setup of three robots.}
	\label{fig:robots1}
\end{figure}

Due to the application, every robot cell must have a complete awareness of its surroundings in order to avoid properly the detected obstacles. Therefore, Moore neighborhood is exploited for every cell to be evolved accordingly (variable $N$ neighborhood radius) with range equal to one. Fig. \ref{fig:robots1} presents such a setup for a team of three robots.

The final variable of the quadruple $F$ has to be defined, meaning the set of the transitions rules, in order to evolve the state of every cell. The applied local transition rules $F:Q^V\rightarrow Q$, considering the applied neighborhood, can be expressed as:

\begin{equation}
\label{eq:EQ5}
C^{t+1}_{x,y}=F(C^t_{x-1,y-1},...,C^t_{x,y},...,C^t_{x+1,y+1})
\end{equation}

or in a more compact, alternative formulation:

\begin{equation}
\label{eq:EQ6}
C^{t+1}_{(x,y)}=F\bigg[\sum_{i=-1}^{i=1} \sum_{j=-1}^{j=1}a_{ij}C^t_{(x+i,y+j)}\bigg]
\end{equation}

\noindent where the $a_{ij}$ are integer constants and thus the function $F$ has a single integer as argument. Due to the discrete nature of the CAs, Cartesian coordinates are mostly utilized to characterize a specific cell. Nonetheless, the orientation of a robot must also be taken into consideration in order to achieve smoother transitions during the evolution. The corresponding parameter must follow the basic principles of a CA meaning the state of orientation should also be finite and integer. An additional parameter, $\theta^t$, is inserted to the transitions rules at each evolution step displaying the following states $\theta^t={0,1,2,3,4}$ which are mapped to the values $\{-90, -45, 0, 45, 90\}$ expressed in degrees. Therefore, equation \ref{eq:EQ6} is transformed as:

\begin{equation}
\label{eq:EQ7}
(C^{t+1}_{(x,y)},\theta^{t+1})=F\bigg[\sum_{i=-1}^{i=1} \sum_{j=-1}^{j=1}a_{ij}C^t_{(x+i,y+j)},\theta^t\bigg]
\end{equation}

The transition set of rules should consider both collision avoidance procedures and the collective behavior of the swarm robot cells. The avoidance of obstacles from a robot cell relies on the characterization of an adjacent cell in its Moore neighborhood as an obstacle cell. During its transit towards the final destination following the shortest path (straight line), the robot cell ``checks'' its contiguous area cell in order to define whether is comprised by free or obstacle cells. If a cell is occupied by an obstacle, the appropriate transition rules will be applied so that it could be bypassed. A small set of the corresponding transition rules are provided in Table \ref{table:table1}. For the frontier cells, null boundary conditions are applied meaning that all virtual cells are always denoted as free cells. 

\begin{table}
	\resizebox{\columnwidth}{!}{%
\begin{tabular}{|c|c|c|c|c|c|c|c|c|c|c|c| }
	
	\hline
\multicolumn{10}{|c|}{Evolution time step t} & \multicolumn{2}{|c|}{Time step t+1} \\
\hline
$C_{(x,y)}$ & $C_{(x-1,y)}$ & $C_{(x-1,y-1)} $& $C_{(x,y-1)}$ & $C_{(x+1,y-1)}$ & $C_{(x+1,y)}$ & $C_{(x+1,y+1)}$ & $C_{(x,y+1)}$ & $C_{(x-1,y+1)}$ & $\theta$ & $C_{(x,y)}$ & $\theta$ \\
\hline
r & 0 & 0 & 0 & 0 & 0 & 0 & 0 & 0 & 2 & f & 2  \\ 
\hline
0 & r & 0 & 0 & 0 & 0 & 0 & 0 & 0 & 2 & r & 2  \\ 
\hline
r & 0 & 0 & 0 & 0 & r+1 & 0 & 0 & 0 & 2 & r & 0  \\ 
\hline
0 & 0 & 0 & r & r+1 & 0 & 0 & 0 & 0 & 0 & r & 0  \\ 
\hline
r & 0 & 0 & 0 & 0 & r+1 & r+1 & r+1 & 0 & 2 & r & 4  \\ 
\hline
0 & 0 & 0 & 0 & 0 & 0 & 0 & 0 & 0 & - & 0 & -  \\ 
\hline
r+1 & 0 & 0 & 0 & 0 & 0 & 0 & 0 & 0 & - & r+1 & -  \\ 
\hline
\end{tabular}
}
\caption{Example of transition rules for collision avoidance.}
\label{table:table1}
\end{table}

In both scenarios (free space or present obstacles), the robot team must display collective behavior as one entity in order to retain or regain their formation. The proposed CA model involves all the appropriate procedures via the application of proper transition rules in order to define the required shortest paths for formation control. It is assumed that robot cells have the ability to exchange data regarding their position in the lattice and in the formation. A local relationship of master and slave is applied between the members of the team. Innermost robot cells are denoted as masters over their neighboring partners while the outermost as slaves. Master robot cells undertake to collect the required information from its slaves and the decide which transition rule should be applied. The latter could be either a command of moving towards one cell to the final destination point or a command of exchanging positions due to a scattered formation. Scattered formations are the result of a detected obstacle so the team members should collaborate to define the shortest paths aiming at recovering their initial structure. Depending on the required application, various transformation can be applied such as straight-line formation, triangular formation etc. The only requirement for the formation control is to define the corresponding CA transition rules where the position in the lattice of every robot cell and additional checks based on their coordinates are required. Table \ref{table:table2} includes a set of such transition rules for straight-line formations where every cell should follow its optimum path (variable $d_i$ denotes the desired path for $i$ robot cell expressed as number of cells).

In case of a deviation, the corresponding robot cell will display a non-zero value between its $d_i$ value and its current vertical coordinate. Thus, the team recognizes that the formation is scattered and the master robot cell apply the necessary transition rules to regain their formation following the extracted shortest paths by exchanging positions in the formation. The extracted optimum paths will eventually lead the team to converge to specific positions in the configuration area and at some point, their formation will be restored. Finally, the team will proceed as one unit towards the final destination by keeping their motion on the desired shortest route. 

\begin{table}
	\resizebox{\columnwidth}{!}{%
		\begin{tabular}{|c|c|c|c|c|c|c|c|c|c|c|c|c| }
			
			\hline
			\multicolumn{11}{|c|}{Evolution time step t} & \multicolumn{2}{|c|}{Time step t+1} \\
			\hline
			Case & $C_{(x,y)}$ & $C_{(x-1,y)}$ & $C_{(x-1,y-1)} $& $C_{(x,y-1)}$ & $C_{(x+1,y-1)}$ & $C_{(x+1,y)}$ & $C_{(x+1,y+1)}$ & $C_{(x,y+1)}$ & $C_{(x-1,y+1)}$ & $\theta$ & $C_{(x,y)}$ & $\theta$ \\
			\hline
			$d_i-y_r=0$ & r & 0 & 0 & 0 & 0 & 0 & 0 & 0 & 0 & 2 & 0 & -  \\ 
			\hline
			$d_i-y_r=0$ & 0 & r & 0 & 0 & 0 & 0 & 0 & 0 & 0 & 2 & r & 2  \\ 
			\hline
			$d_i-y_r > 0$ & r & 0 & 0 & 0 & 0 & 0 & 0 & 0 & 0 & 2 & 0 & -  \\ 
			\hline
				$d_i-y_r > 0$ & 0 & 0 & 0 & 0 & r & 0 & 0 & 0 & 0 & 2 & r & 2  \\ 
			\hline
				$d_i-y_r < 0$ & r & 0 & 0 & 0 & 0 & 0 & 0 & 0 & 0 & 2 & 0 & -  \\ 
			\hline
			$d_i-y_r < 0$ & 0 & 0 & 0 & 0 & 0 & 0 & r & 0 & 0 & 2 & r & 2  \\ 
			\hline
		\end{tabular}
	}
	\caption{Example of transition rules for formation control over a straight line.}
	\label{table:table2}
\end{table}

\subsection{Implementation in Real Swarm Robot Team}
\label{subsec:implementationRobots}

\begin{figure}[h!]
	\centering
	\includegraphics[scale=0.7]{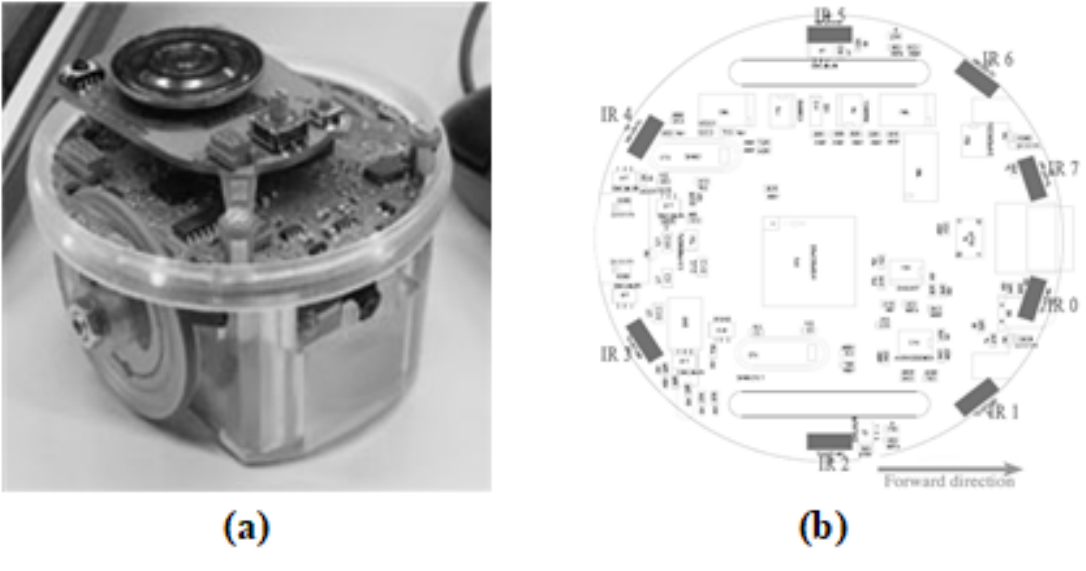}
	\caption{(a) E-puck robot and (b) IR proximity sensors.}
	\label{fig:epuck}
\end{figure}

The main objective of the method is to display low computational and memory requirements so that it could be developed as a firmware and loaded on real robots. The simplicity of the developed CA renders the method suitable to achieve this task. To be fully functional, every utilized robot must be equipped with a proper hardware architecture including a microprocessor, distance sensors (e.g. IR), step motors and a communication interface (e.g. Bluetooth). No central control (e.g. base station) is required following the basic principles of the swarm robotics theory. Each robot of the team should be loaded with the method's implementation in order for the system to accomplish its goals. The team could also include different types of robots, forming a heterogeneous swarm, with the only restriction that all the robots exploit the same communication protocol. For testing purposes, without loss of generality, the method was tested on a three member's squad of miniature robots, called E-puck \cite{mondada2009puck} (Fig. \ref{fig:epuck}(a)). The e-puck robotic architecture comprises a fully open source platform providing full access to every of its modules. Equipped with all the aforementioned requirements for the method's implementation, it is a valid selection for swarm robotics applications.

The first stage of implementation is the determination of the lattice size that is proportional to the desired distance to be covered and the cell length. More specifically, cell length is strictly related to the proximity sensors' readings. In order to identify the presence of an obstacle, the sensors are enabled and based on the acquired data; an adjacent cell (which corresponds to an actual fragment of space) is denoted as free or as an obstacle cell. The IR sensors mounted on the e-puck (Fig.\ref{fig:epuck}(b)) produce a scalable number, which represents the distance from an object. A higher value corresponds to a lower distance of an object and vice versa. Nonetheless, due to the nature of these sensors, their response is affected by the environment's ambient light leading to false positives (object detections) in large distances. On the other hand, in case of small cell lengths, the lattice is increased leading to higher memory resources and less accuracy in denoting a cell as obstacle due to the placement of the sensors on the robot. Therefore, multiple experiments were conducted in order to identify the proper cell size and ensure the required accuracy of the sensor's readings.

\begin{figure}[h!]
	\centering
	\includegraphics[scale=0.5]{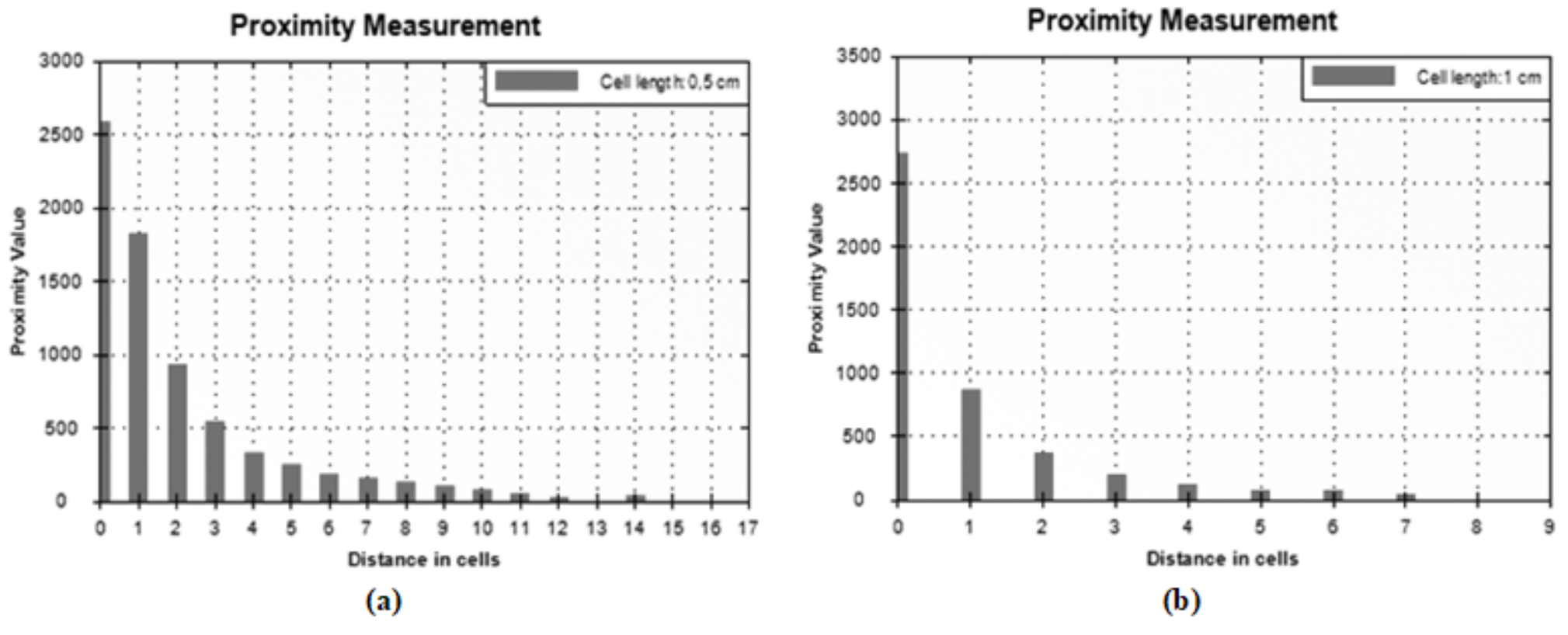}
	\caption{Measures of the IR proximity sensors for different lengths: (a) 0.5 cm and, (b) 1 cm.}
	\label{fig:IR}
\end{figure}

For this task, a special software was developed on a personal computer to help us acquire all the sensors' data and model their response. The software was connected with an e-puck robot via Bluetooth in order to acquire the required readings. At first, the smallest possible distance between the robot and an object was applied. At every time step, a backward motion was applied, covering a distance of one cell length and capturing the response of the sensor. When the maximum possible proximity (8cm) was covered, the front sensors' (IR7 and IR0 of Fig. \ref{fig:epuck}(b)) responses were transmitted back to the software for visualization and evaluation purposes. Multiple cell sizes were tested while Fig. \ref{fig:IR} includes the sensors' responses for two cell lengths, 0.5cm and 1cm.

\begin{figure}[h!]
	\centering
	\includegraphics[scale=0.7]{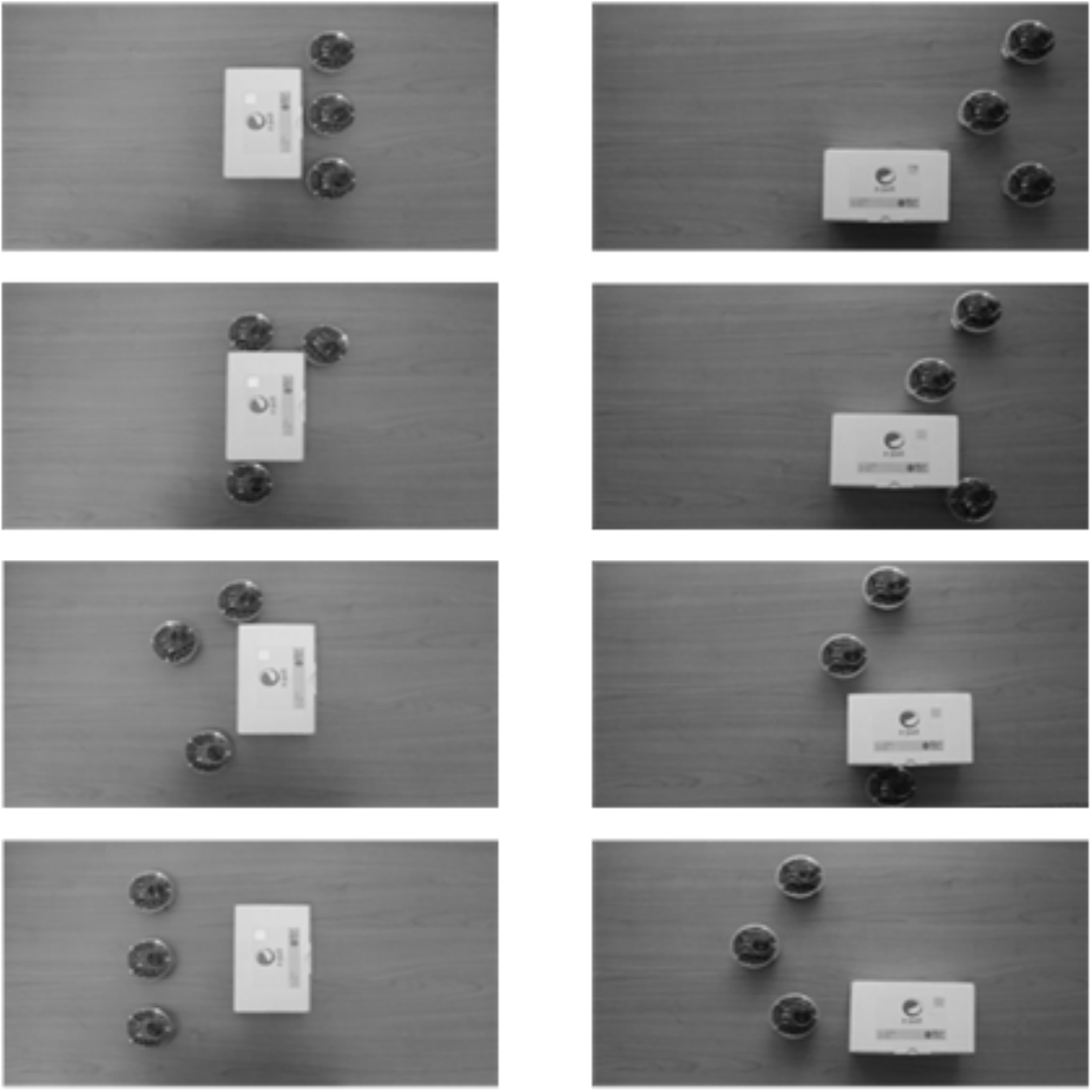}
	\caption{Swarm robots from top to bottom in (a) straight line formation and, (b) triangular formation.}
	\label{fig:moves}
\end{figure}

All actions executed by every robot can be summarized into two different subsets of execution. During the first stage, every robot acts as an individual and ``scans'' its adjacent environment in order to detect potential obstacles. The IR proximity sensors are enabled and according to their readings, the corresponding cells are denoted as free or obstacle cells. The presence of an obstacle is detected by comparing the acquired scalable values with the value that represents the cell length. For the tested environment, a cell length equal to 0.5 cm was used. In addition, two different formations were tested, namely a straight-line and a triangular formation. Snapshots of the entire procedure are provided in Fig. \ref{fig:moves}.

More specific, for the straight-line formation, all robots are deployed to their initial positions forming a straight line. As they cover the desired distance towards, their final destination, all robots detect an obstacle that interfere their motion. Thus, obstacle avoidance transition rules are applied to bypass the obstacle. At the second stage, the central robot, which acts as a master, commands its right adjacent robot to exchange their positions in order to recover the team's formation. Collective behaviors are executed and following the extracted shortest paths, the formation is retrieved with the minimum time cost. Finally, the team continues to their final positions covering the defined distance.

Similar process is followed for the triangular formation. All robots of the team are initially deployed forming a triangular formation. At some point, the left most robot detects an obstacle that must be avoided. The appropriate transition rules are applied to achieve this task. During that process, the central/master robot coordinates all the motions with the rest of the team members and decides that no position shifting is required. Until the robot avoids the box, the rest of the team freezes its motions and wait the discarded robot to regain its position to the formation. For this type of formation control, both vertical and horizontal coordinates in the lattice of every robot are exchanged. Since the formation is recovered, the team proceeds to its final destination as one entity, again via collaborations.

\section{Physarum Polycephalum CA model}
\label{sec:physarumModel}

Modern computers offer sufficient processing power to handle most of the analysis that several complex phenomena require. Physics, biology or chemistry can be characterized as complex phenomena. They are based on processes and systems using inhomogeneities, multiple interactions and complex constraints that even the modern computers cannot handle. CAs include all the necessary characteristics (handling of complex boundary and initial conditions, description of local interaction of a system with inhomogeneities and anisotropies that lead to global behavior, inherent parallelism) that makes them the appropriate tool to model and simulate natural phenomena. 

A fungus, Physarum polycephalum, \index{Physarum polycephalum} is such a system. Physarum polycephalum is a large amoeba-like cell consisting of a dendritic network of tube-like structures (pseudopodia). It changes its shape as it crawls over a plain agar gel, and if nutrients is placed at two different points, it will extend pseudopodia that connect the two nutrient sources (FSs). Nakagaki \emph{et al.} \cite{nakagaki2000intelligence} showed that this simple organism has the ability to find the minimum-length solution between two points in a labyrinth. This resulted in an intensive period of research on this organism that exposed a great range of its computational abilities to spatial representations of various graph problems. CAs are used extensively in this system because they have the ability to model the foraging behavior of plasmodium (physarum in its nutritious stage). Plasmodium spreads its pseudopodia and searches for chemo-attractants to lead it to nutrients that can devour and survive. It is very important for the survival of this life form to consume the least possible energy to find this chemo-attractants. This is the reason why the plasmodium creates tubes with minimum distance between food spots in a maze. 

CAs is the most suitable paradigm to model such a structure \cite{Tsompanas2012,dourvas2015hardware,tsompanas2015slime,evangelidis2015slime,Tsompanas15,Tsompanas15plus,Kalogeiton2015,Kalogeiton15,Tsompanas2016,Dourvas2016,Tsompanas16}. The maze can be modeled by creating a grid of cells with standard initial and boundary conditions. Plasmodium is not a unified mass but it is composed by many elementary parts that communicate and move. This local interaction leads in the movement of the whole plasmodium's mass. Each CA cell can model this elementary part of Physarum. The neighborhood of this cell will include walls of the maze, empty paths or other plasmodium's particles. This cell will interact with its environment, exchange stimuli and information and finally it will take a decision about the next direction of its movement. The evolution of this CA system leads to the final solution of the maze. 

\begin{figure}[htbp]
	\centering
	\includegraphics[scale=0.2]{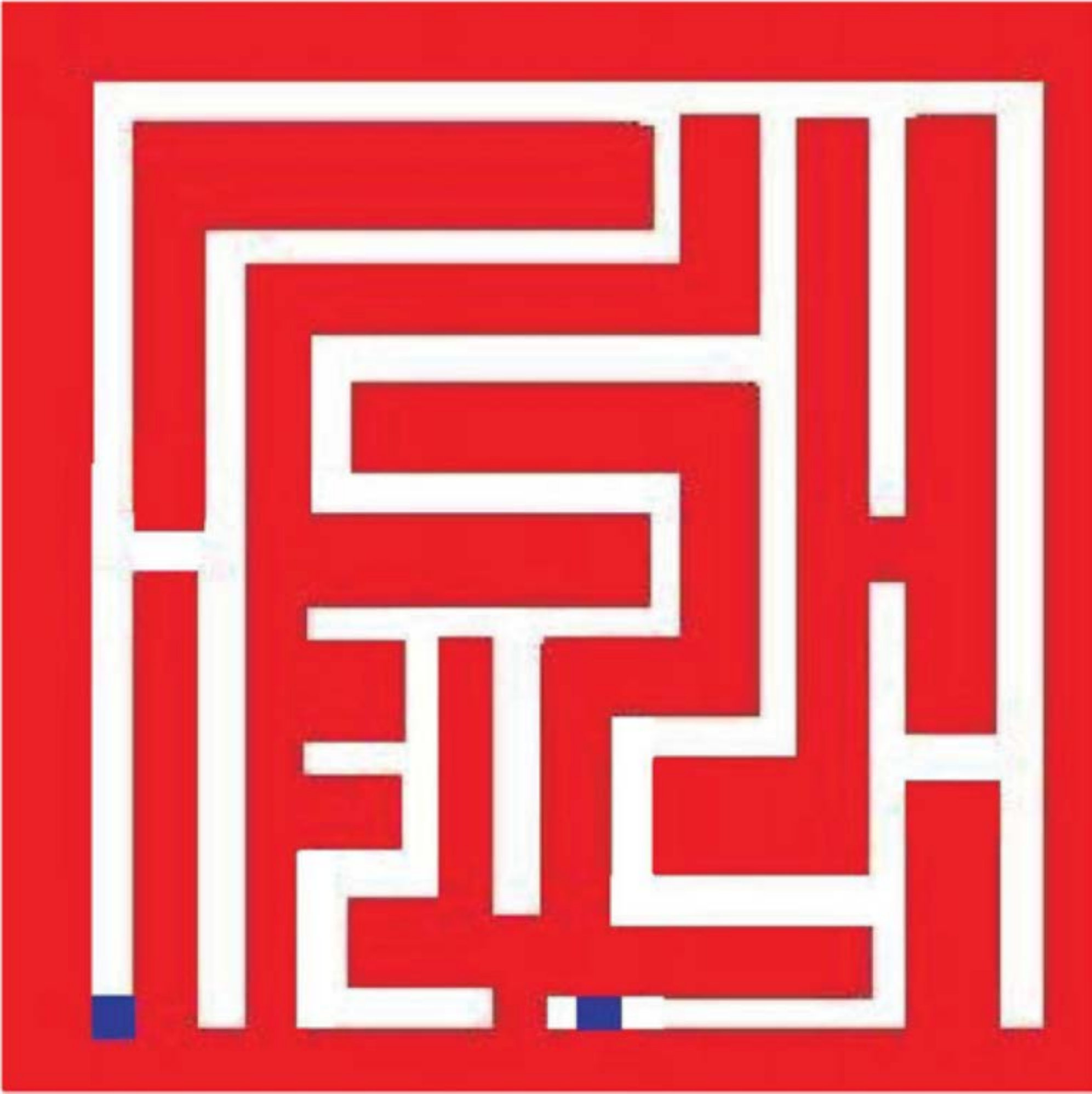}
	\caption{The under study maze in correspondence to the one of \cite{nakagaki2000intelligence}.}
	\label{fig:antiFig1}
\end{figure}

The maze used for the biological experiment (Fig. \ref{fig:antiFig1}) was also used as an input for our algorithm. More specifically, Nakagaki \emph{et al.} \cite{nakagaki2000intelligence} took a growing tip of an appropriate size from a large plasmodium in a 25 $\times$ 35 $cm^2$
culture trough and divided it into small pieces.

\begin{figure}[htbp]
	\centering
	\includegraphics[scale=0.3]{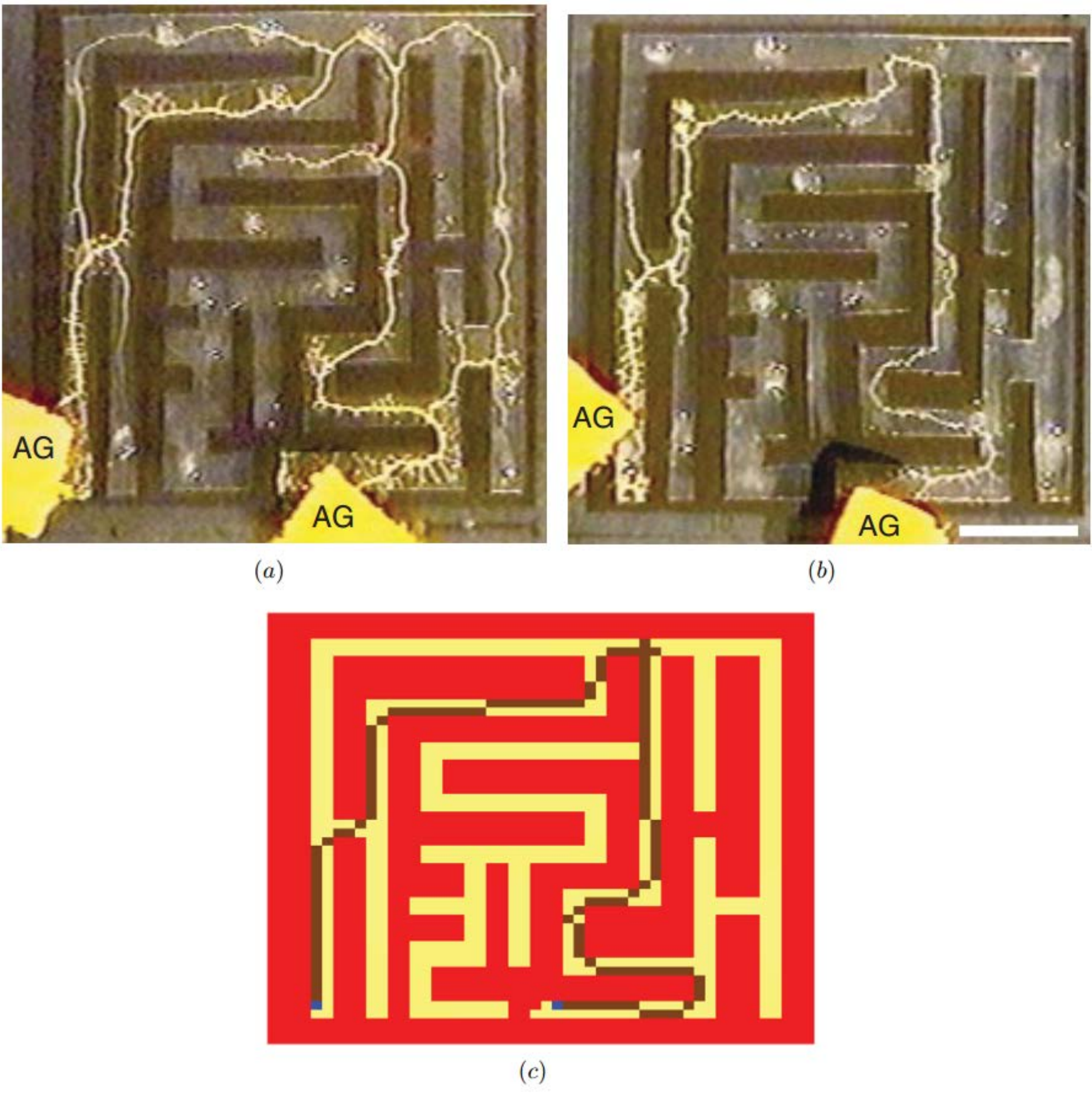}
	\caption{(a), (b) The maze solving by P. polycephalum after 4 and 8 hours, respectively, as presented by Nakagaki \emph{et al.} in \cite{nakagaki2000intelligence}, where yellow color, is the plasmodium and black are the maze `walls'. (c) The final simulation results, after 1500 time steps, of the CA that mimics the P. Polycephalum's behavior for the same maze. The situation in (b) is successfully reproduced.}
	\label{fig:antiFig3}
\end{figure}

Then, they positioned these in a maze created by cutting a plastic film and placing it on an agar surface. The plasmodial pieces spread and coalesced to form a single organism that filled the maze, avoiding the dry surface of the plastic film. At the start and end points of the maze, they placed agar blocks containing nutrient (ground oat flakes) and there were four possible routes between the start and the end points. The plasmodium pseudopodia reaching dead ends in the labyrinth shrank (Fig. \ref{fig:antiFig3}(a)), resulting in the formation of a single thick pseudopodium spanning the minimum length between the nutrient-containing agar blocks (Fig. \ref{fig:antiFig3}(b)). In our case, we artificially reconstructed the aforementioned maze taking into consideration the exact positions of the maze. 

\begin{figure}[htbp]
	\centering
	\includegraphics[scale=0.3]{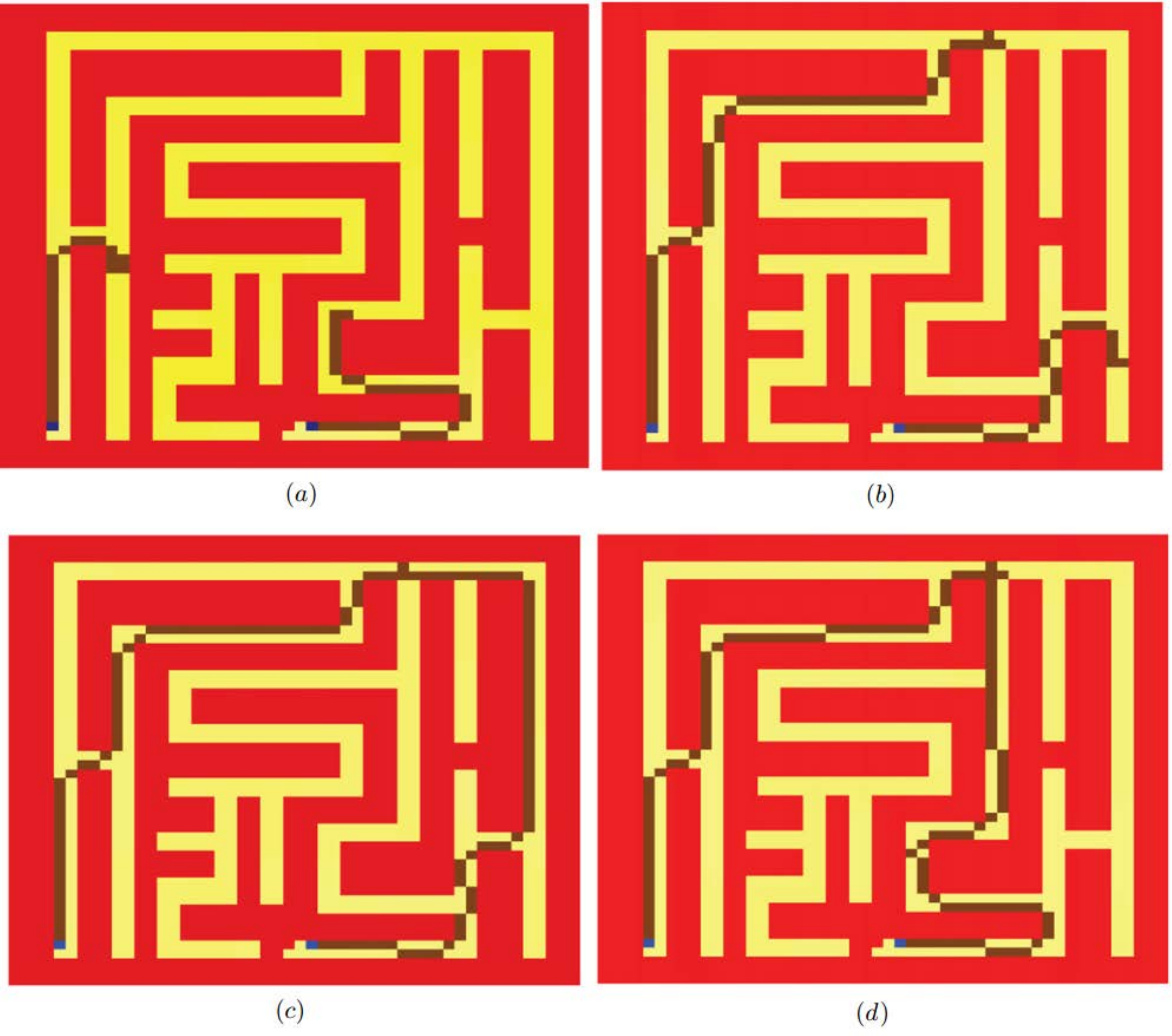}
	\caption{The amoeba-like CA simulation results for the maze after (a) 500, (b) 1000, (c) 1200 and (d) 1500 time steps,
		respectively.}
	\label{fig:antiFig2}
\end{figure}

In order to simulate this biological experiment, the area is divided into a matrix of squares with identical areas and each square of the surface is represented by a CA cell. The type of neighborhood that was used in this CA model is the Moore neighborhood which means that we use the north, south, east, west, north-east, north-west, south-east and south-west neighbors.The state of the $(i,j)$ cell at time $t$, defined as $C^t_{i,j}$ is equal to:

\begin{equation}
C^t_{i,j}=\{Topology_{i,j}, Chem^t_{i,j}, Dir^t_{i,j}, Phys^t_{i,j}, Pseudo^t_{i,j}\}
\label{eq1}
\end{equation}

\begin {itemize}

\item
$Topology_{i,j}$ is a variable which indicates the type of area of the corresponding $(i,j)$ cell. The possible values of this variable are 0,1,2,3 and indicate a free area, the spot of the initially placed FS, the spot of the initially placed plasmodium and the spot which represents a wall of the topology respectively.

\item
$Chem^t_{i,j}$ represents the concentration of chemo-attractants at time $t$ in the area corresponding to the $(i,j)$ cell. In order to calculate this variable for every cell, we make use of the concentration of the neighborhood to update the value of the central cell.

\item
$Dir^t_{i,j}$ is a variable that indicates the direction of the attraction of the plasmodium by the chemicals produced by the FS. For example, if the area around a corresponding cell has no chemo-attractants, the foraging strategy of the plasmodium is uniform and, thus, these parameters are equal to zero. If there is higher concentration of chemo-attractants in the cell at direction $x$ from the one in direction $y$, then the parameter corresponding to direction $x$ is positive and the parameter corresponding in the direction $y$ is negative. This happens, in order to more accurately simulate the non-uniform foraging behavior of the plasmodium.

\item
$Phys^t_{i,j}$ indicates the volume of the cytoplasmic material of the plasmodium in the corresponding $(i,j)$ cell. In order to calculate this variable for every cell, we make use of the neighbor's volumes.

\item
Finally, $Pseudo^t_{i,j}$ is a variable which can take values [0,1] and illustrates if the $(i,j)$ cell is included in the final path of tubular network that is formed inside the plasmodium's body. This tubular network forms the shortest path between the FSs and the cell from where the plasmodium started to expand and it is our final solution.
\end {itemize}

The amoeba-like CA model simulation results after 500, 1000, 1200 and 1500 time steps are shown in Fig. \ref{fig:antiFig2}. Compared to the results of the biological experiment, which are presented in Fig. \ref{fig:antiFig3}, the algorithm can be considered successful. As is illustrated in Fig. \ref{fig:antiFig2}, it takes 1200 time steps to find a solution that is not the best one. However, after 1500 time steps, it manages to solve the maze using the shortest possible route. It should be noted that in analogy to the real experiments, the amoeba-like CA model changes its shape in the maze to form one thick tube covering the shortest distance between the FSs, so as to maximize its foraging efficiency, and therefore, its chances of survival. 

The period of 1500 time steps (which correspond to about 45 s of real time on a PC) may seem like a long time, but compared to 8 h needed for the biological experiment, it is not a significantly long time period.

\subsection{GPGPU implementation}
\label{sec:gpuImplem}

The term GPGPU (General-Purpose computing on Graphics Processing Units) refers to the use of the GPU processor as a parallel device for purposes other than graphic elaboration. More specific, GPU is often used in order to solve some complex computational problems that classical CPU cannot handle. 
This device has the ability to execute a great number of independent threads in parallel. So if a complex problem has an inherent parallel nature, an implementation in GPU is going to multiply the performance of its algorithm and the solution will be produced much faster. The GPU's architecture has a computational power that can exceed a teraFLOP and it is fully suitable for fine grain parallelism.% 
The reason of the great success and enormous spread of the GPGPU application in the past few years, is CUDA programming model. %\cite{nvidia}. 
The basic structure of CUDA is that it provides three key abstractions, namely the hierarchy with which the threads are organized, the memory organization and the functions that execute in parallel, called kernels.

 In a CUDA application, some parts are performed in a parallel way and some other parts are performed in the classical serial way. The $device$, which is the name of the GPU in CUDA context, can be thought as an additional co-processor of the main CPU which is called $host$ in the CUDA context. In order to take off the performance of our algorithm, we have to exploit the parts of the data that are made to work in parallel and execute them on the device as many time steps is necessary. In order to achieve this, we have to call one, two or more kernels which use thousands of threads.

One key problem towards the implementation in a GPU is the way the memory is organized and used. The threads can be organized and cooperate together by sharing a common fast $shared-memory$ synchronizing in some points of the kernel within a so called $thread-block$. But the number of threads that a block can use is limited and for most of the applications more parallelism is needed. CUDA gives the choice to launch kernels with a larger total number of threads by organizing block of threads together, by means of a grid of blocks. So it is possible to choose a cell of a CA, which can be thought as a particular data of the device memory, and associate it to a current thread of a kernel.

Threads can access different memory locations during execution. Generally, there are three types of memory used in CUDA applications, namely (\textit{a}) the private memory, which is the memory its thread has for its own, (\textit{b}) the shared memory, which is the memory being visible to all threads in a block and ({c}) the global memory, which is a larger memory on the device board but it is outside the computing chip.
In this study, we make use of the global memory of the device. This memory is slower if compared with the shared memory but it can deliver a significantly higher memory bandwidth than the traditional CPU memory. It is measured that is about 20 times more efficient to access the global memory of the GPU than the CPU memory. As a result, when a CUDA application is designed, the minimum data transfers between CPU and GPU should take place.

The reason why GPGPU programming is used for CAs models can be explained easily when referring to the CAs' parallel nature. The local interaction of the neighbors that CAs methods propose is another fact that makes these implementations very suitable and very fast. These features make the CAs models ideal to be implemented in parallel computers. The basic idea when computing a CA model in GPU, which is also used in our implementation, can be described as follows: First, we compute the next state of all the cells in parallel. Afterward we use two memory regions to store the data. More specifically, we use one region for the $CA_{current}$, which indicates the CA states before the calculations and one for the $CA_{next}$, which in turns indicates the CA states after the calculations. Finally, the switching between the $CA_{current}$ and the $CA_{next}$ in each time step takes place.

For this paper we store the CA data to the global memory of the device. The steps of the algorithm are: (\textit{a}) Split the CA states and make use of a kernel for every one of them. In more detail, we make use of a kernel to hold the $Topology_{i,j}$, one kernel for the computation of the diffusion equation of the chemo-attractants, $Chem_{i,j}$, and their direction, $Dir_{i,j}$, one kernel for the computation of the diffusion equation of the mass of plasmodium, $Phys_{i,j}$ and finally one kernel that computes the $Pseudo_{i,j}$ to find the shortest path in the maze. (\textit{b}) An initialization of the current state for all these kernels happens through a CPU-GPU memory copy operation (i.e. from host to the device global memory). (\textit{c}) Every kernel runs in each time step and makes its calculations by using the information of the states of the other necessary kernels. For example, in order to calculate the $Pseudo_{i,j}$ we have to know which of the neighbors has the greater mass value. Therefore, we take this information from the kernel that executes the computation of the $Phys_{i,j}$. (\textit{d}) At the end of each CA step, a device to device memory copy operation is used to update the new values in order to continue the calculations in the next CA step. (\textit{e}) When the simulation is completed the final state of the automaton is being retrieved from the global memory of the device to the host through a GPU-CPU memory copy operation.

For the proposed GPU implementation of the presented slime mould CA based model we used the graphics card NVIDIA GT640.  We used a $50 \times 50$ CA cells in order to synthesize the maze.  The time needed for the presented solution to result is 2.47 seconds. In the serial code the time needed for the serial software implementation in MATLAB was approximately 45 seconds. Therefore, the increase in the performance in our implementation is about 18.2 times more than the one in MATLAB.

\subsection{Hardware implementation}
\label{subsec:hardware}

Current FPGAs include logic density equivalent to millions of gates per chip and can implement very complex computations. CAs consist of a uniform $n$-dimensional structure, composed of many identical synchronous cells where both memory and computation are involved, thus matching the inherent design layout of FPGA Hardware. As a result memory and processing unit are closely related both in CAs cells and FPGA configurable logic blocks (CLBs). The structure of a cell consists of a combinational part connected with one or more memory elements in a feedback loop shape while the state of the memory elements is also defined by the inputs and the present state of these elements. For this implementation the design produced by using VHDL code has been analyzed and synthesized by Quartus II (32-bit version 12.1 build) FPGA design software of ALTERA Corporation.

Each CA cell is implemented by a hardware block called ``PhysarumCell''. Each ``PhysarumCell'' block is connected appropriately with its four neighbors (west, east, south and north). It uses the inputs from the neighbors and the previous state of itself to produce results that simulate the movement of the plasmodium. A ``PhysarumCell'' block has 22 inputs and 7 outputs. 

After creating the lattice, the user has to provide only the topology of the experimental area %maze 
by giving values to the 2-bit signals $topology$ for each individual cell, namely the location of the FS and the location of the initial introduction of the plasmodium to the experimental area and the parameters for the diffusion equations.

\begin{table}
	\centering
	\begin{tabular}{l c }\\
		\hline
		%%%{} &{} &{} &{} &{} &{}\\[-9pt]
		%%%{} &{} &3 &4 &8 &10\\
		%%%\hline
		{} &{} \\[-9pt]
		Quartus II 32-bit Version & 12.1 Build 243 01/31/2013 SP 1\\
		Total logic elements & 1,739\\
		Total registers & 45\\
		Total pins & 226\\
		\hline\\
	\end{tabular}
\caption{FPGA hardware implementation details for one CA cell.}
\label{taCAcell}
\end{table}

The number of logic elements, registers and pins of the CA cell are presented in the above Table \ref{taCAcell}. Moreover, to illustrate the area needed for a fully interconnected system of a CA grid implementing the proposed bio-inspired model, the results of synthesizing a $10\times{10}$, a $15\times{15}$ and a $20\times{20}$ grid are illustrated in Table \ref{tabtotalRegisters}. The circuits are synthesized on several target devices and the results on the Stratix V 5SGXBB are presented here. The process ends in a few $\mu$s. In Table \ref{tabtotalRegisters} below, it is shown that for almost every 150 CA cells there is an increment of around 300,000 logic elements on average.

%----------------------------------------------------------------------------------------------------
\begin{table}
	\centering
	\begin{tabular}{l c c c}\\
		\hline
		{} &{} &{} &{} \\[-9pt]
		{} &$10\times{10}$ &$15\times{15}$ &$20\times{20}$\\
		\hline
		{} &{} &{} &{}\\[-9pt]
		Total logic elements &161,162 &370,447 &666,060\\
		Total registers &8,360 &18,840 & 33,520\\
		Total pins & 317 & 692 & 1217\\
		\hline\\
		\label{table:totalRegisters}
	\end{tabular}
	\caption{FPGA hardware implementation for different topology sizes.}
	\label{tabtotalRegisters}
\end{table}

\section{Conclusions}
\label{conclusions}
In this chapter the inseparable relationship between CAs and shortest path problem is depicted. CAs are a very powerful modeling tool that can capture the essential characteristics of this problem and produce effective results. They can manage the classical $S^3P$, $S^3DSP$ and $APSP$ problems as presented in section \ref{sec:adamatzky}. They can learn and find solutions in a stochastic graph as presented in section \ref{sec:DLA}. They can also move to three-dimensional space and provide solutions to difficult territories as already shown in corresponding subsections. But their use is not only theoretical. They can be applied successfully for computer networks main problems as shown in section \ref{sec:net}. They are a very useful tool in robotics and their movement inside a maze section as demonstrated in section \ref{sec:swarmRobots}. They can also describe and model very effectively and efficiently physical phenomena and living structures that have the ability to provide unconventional solution to shortest path problems as presented in section \ref{sec:physarumModel}. What is the reason that the complexity bounds are so good? It takes place because we used a very restricted form of a rectangular lattice to make the structure of a problem most resembling the architecture of computing device which solves this problem. Can CAs be applied in practice? Of course. The CAs algorithms are derived almost directly for biology and nature. For this reason, the implementation of these algorithms in massively parallel processors or neurocomputers is an event that already happens.

%%%\section{References}
%\bibliographystyle{plain}
%\bibliography{sirakoulis/mybibfile}

%\bibliographystyle{spmpsci}
%\bibliography{mybibfile2}

\begin{thebibliography}{10}
\providecommand{\url}[1]{{#1}}
\providecommand{\urlprefix}{URL }
\expandafter\ifx\csname urlstyle\endcsname\relax
  \providecommand{\doi}[1]{DOI~\discretionary{}{}{}#1}\else
  \providecommand{\doi}{DOI~\discretionary{}{}{}\begingroup
  \urlstyle{rm}\Url}\fi

\bibitem{adamatzky1994}
Adamatzky, A.: Identification Of Cellular Automata.
\newblock Taylor \& Francis (1994)

\bibitem{adamatzky1996computation}
Adamatzky, A.: Computation of shortest path in cellular automata.
\newblock Mathematical and Computer Modelling \textbf{23}(4), 105--113 (1996)

\bibitem{Beigy}
Beigy, H., Meybodi, M.R.: Utilizing distributed learning automata to solve
  stochastic shortest path problems.
\newblock International Journal of Uncertainty, Fuzziness and Knowledge-Based
  Systems \textbf{14}(05), 591--615 (2006)

\bibitem{PACT95}
C., H., Hoffmann, R., S., W.: Compilation of cdl for different target
  architecures.
\newblock In: V.~Malyshkin (ed.) Parallel Computing Technologies, pp. 169--179
  (1995)

\bibitem{MPCS96a}
C., H., R., H.: Cdl --- a language for cellular processing.
\newblock In: G.~Sechi (ed.) Proceedings of the Second International Conference
  on Massively Parallel Computing Systems, pp. 41--64 (1996)

\bibitem{ACRI96}
C., H., R., H.: Solving routing problems with cellular automata.
\newblock In: Proceedings of the Second Conference on Cellular Automata for
  Research and Industry (ACRI '96), pp. 89--98 (1996)

\bibitem{charalampous2012efficient}
Charalampous, K., Amanatiadis, A., Gasteratos, A.: Efficient robot path
  planning in the presence of dynamically expanding obstacles.
\newblock Cellular Automata pp. 330--339 (2012)

\bibitem{charalampous2014real}
Charalampous, K., Kostavelis, I., Amanatiadis, A., Gasteratos, A.: Real-time
  robot path planning for dynamic obstacle avoidance.
\newblock Journal of Cellular Automata \textbf{9} (2014)

\bibitem{defoort2008sliding}
Defoort, M., Floquet, T., Kokosy, A., Perruquetti, W.: Sliding-mode formation
  control for cooperative autonomous mobile robots.
\newblock IEEE Transactions on Industrial Electronics \textbf{55}(11),
  3944--3953 (2008)

\bibitem{Dijkstra}
Dijkstra, E.W.: A note on two problems in connexion with graphs.
\newblock Numer. Math. \textbf{1}, 269--271 (1959)

\bibitem{dourvas2015hardware}
Dourvas, N., Tsompanas, M.A., Sirakoulis, G.C., Tsalides, P.: Hardware
  acceleration of cellular automata physarum polycephalum model.
\newblock Parallel Processing Letters \textbf{25}(01), 1540,006 (2015)

\bibitem{Dourvas}
Dourvas, N.I., Sirakoulis, G.C., Adamatzky, A.: Cellular automaton
  belousov�zhabotinsky model for binary full adder.
\newblock International Journal of Bifurcation and Chaos \textbf{27}(06),
  1750,089 (2017).
\newblock \doi{10.1142/S0218127417500894}

\bibitem{Dourvas2016}
Dourvas, N.I., Tsompanas, M.A.I., Sirakoulis, G.C.: Parallel Acceleration of
  Slime Mould Discrete Models, pp. 595--617.
\newblock Springer International Publishing, Cham (2016)

\bibitem{evangelidis2015slime}
Evangelidis, V., Tsompanas, M.A., Sirakoulis, G.C., Adamatzky, A.: Slime mould
  imitates development of roman roads in the balkans.
\newblock Journal of Archaeological Science: Reports \textbf{2}, 264--281
  (2015)

\bibitem{ferguson2006using}
Ferguson, D., Stentz, A.: Using interpolation to improve path planning: The
  field d* algorithm.
\newblock Journal of Field Robotics \textbf{23}(2), 79--101 (2006)

\bibitem{Floyd}
Floyd, R.W.: Algorithm 97: Shortest path.
\newblock Communications of the ACM \textbf{5}(6), 345 (1962)

\bibitem{garcia2009path}
Garcia, M.P., Montiel, O., Castillo, O., Sep{\'u}lveda, R., Melin, P.: Path
  planning for autonomous mobile robot navigation with ant colony optimization
  and fuzzy cost function evaluation.
\newblock Applied Soft Computing \textbf{9}(3), 1102--1110 (2009)

\bibitem{Georgoudas2010}
Georgoudas, I.G., Koltsidas, G., Sirakoulis, G.C., Andreadis, I.T.: A Cellular
  Automaton Model for Crowd Evacuation and Its Auto-Defined Obstacle Avoidance
  Attribute, pp. 455--464.
\newblock Springer Berlin Heidelberg, Berlin, Heidelberg (2010)

\bibitem{Giitsidis}
Giitsidis, T., Sirakoulis, G.C.: Modeling passengers boarding in aircraft using
  cellular automata.
\newblock IEEE/CAA Journal of Automatica Sinica \textbf{3}(4), 365--384 (2016)

\bibitem{HUSSAIN}
H., H.: Integration eines Compilers fur die Zellularsprache CDL in das
  XCellsim--System.
\newblock Techn. Univ. Darmstadt, Comp. Science Dept. (1994)

\bibitem{hwu2017self}
Hwu, T., Isbell, J., Oros, N., Krichmar, J.: A self-driving robot using deep
  convolutional neural networks on neuromorphic hardware.
\newblock In: Neural Networks (IJCNN), 2017 International Joint Conference on,
  pp. 635--641. IEEE (2017)

\bibitem{IoannidisApp}
Ioannidis, K., Sirakoulis, G.C., Andreadis, I.: {A path planning method based
  on Cellular Automata for Cooperative Robots}.
\newblock Applied Artificial Intelligence \textbf{25}(8), 721--745 (2011)

\bibitem{IoannidisR}
Ioannidis, K., Sirakoulis, G.C., Andreadis, I.: {Cellular Ants: A Method to
  Create Collision Free Trajectories for a Cooperative Robot Team}.
\newblock Robotics and Autonomous Systems \textbf{59}(2), 113--237 (2011)

\bibitem{IoannidisJCA}
Ioannidis, K., Sirakoulis, G.C., Andreadis, I.: {Cellular Automata-based
  Architecture for Cooperative Miniature Robots}.
\newblock Journal of Cellular Automata \textbf{8}(1-2), 91--111 (2013)

\bibitem{Li2006}
J., L., B.H., W., P.Q., J., T., Z., W.X., W.: Growing complex network model
  with acceleratingly increasing number of nodes.
\newblock Acta Physica Sinica \textbf{55}(8), 4051--4057 (2006)

\bibitem{Johnson}
Johnson, D.B.: A note on dijkstra's shortest path algorithm.
\newblock J. ACM \textbf{20}(3), 385--388 (1973)

\bibitem{Kalogeiton15}
Kalogeiton, V., Papadopoulos, D., Georgilas, I., Sirakoulis, G., Adamatzky, A.:
  Cellular automaton model of crowd evacuation inspired by slime mould.
\newblock International Journal of General Systems \textbf{44}(3), 354--391
  (2015)

\bibitem{Kalogeiton2015}
Kalogeiton, V.S., Papadopoulos, D.P., Georgilas, I.P., Sirakoulis, G.C.,
  Adamatzky, A.I.: Biomimicry of Crowd Evacuation with a Slime Mould Cellular
  Automaton Model, pp. 123--151.
\newblock Springer International Publishing, Cham (2015)

\bibitem{Kechaidou}
Kechaidou, M., Sirakoulis, G.: Game of life variations for image scrambling.
\newblock Journal of Computational Science \textbf{21}(Supplement C), 432 --
  447 (2017)

\bibitem{Konsta}
Konstantinidis, K., Amanatiadis, A., Chatzichristofis, S.A., Sandaltzopoulos,
  R., Sirakoulis, G.C.: Identification and retrieval of dna genomes using
  binary image representations produced by cellular automata.
\newblock In: 2014 IEEE International Conference on Imaging Systems and
  Techniques (IST) Proceedings, pp. 134--137 (2014)

\bibitem{lee}
Lee, C.Y.: An algorithm for path connections and its applications.
\newblock IRE Transactions on Electronic Computers \textbf{EC-10}(2), 346--365
  (1961)

\bibitem{liang2015efficient}
Liang, J.H., Lee, C.H.: Efficient collision-free path-planning of multiple
  mobile robots system using efficient artificial bee colony algorithm.
\newblock Advances in Engineering Software \textbf{79}, 47--56 (2015)

\bibitem{liu2014dynamic}
Liu, S., Sun, D., Zhu, C.: A dynamic priority based path planning for
  cooperation of multiple mobile robots in formation forming.
\newblock Robotics and Computer-Integrated Manufacturing \textbf{30}(6),
  589--596 (2014)

\bibitem{macwan2015multirobot}
Macwan, A., Vilela, J., Nejat, G., Benhabib, B.: A multirobot path-planning
  strategy for autonomous wilderness search and rescue.
\newblock IEEE transactions on cybernetics \textbf{45}(9), 1784--1797 (2015)

\bibitem{marchese2015multi}
Marchese, F.M.: Multi-resolution hierarchical motion planner for multi-robot
  systems on spatiotemporal cellular automata.
\newblock In: Robots and Lattice Automata, pp. 149--173. Springer (2015)

\bibitem{Mardiris}
Mardiris, V.A., Sirakoulis, G.C., Karafyllidis, I.G.: Automated design
  architecture for 1-d cellular automata using quantum cellular automata.
\newblock IEEE Transactions on Computers \textbf{64}(9), 2476--2489 (2015)

\bibitem{mastellone2007remote}
Mastellone, S., Stipanovic, D.M., Spong, M.W.: Remote formation control and
  collision avoidance for multi-agent nonholonomic systems.
\newblock In: Robotics and Automation, 2007 IEEE International Conference on,
  pp. 1062--1067. IEEE (2007)

\bibitem{Wang2012}
Min~Wang Yongsheng~Qian, X.G.: Improved calculation method of shortest path
  with cellular automata model.
\newblock Kybernetes \textbf{41}(3-4), 508--517 (2012)

\bibitem{moghaddam2016planning}
Moghaddam, S.K., Masehian, E.: Planning robot navigation among movable
  obstacles (namo) through a recursive approach.
\newblock Journal of Intelligent \& Robotic Systems \textbf{83}(3-4), 603--634
  (2016)

\bibitem{mondada2009puck}
Mondada, F., Bonani, M., Raemy, X., Pugh, J., Cianci, C., Klaptocz, A.,
  Magnenat, S., Zufferey, J.C., Floreano, D., Martinoli, A.: The e-puck, a
  robot designed for education in engineering.
\newblock In: Proceedings of the 9th conference on autonomous robot systems and
  competitions, vol.~1, pp. 59--65. IPCB: Instituto Polit{\'e}cnico de Castelo
  Branco (2009)

\bibitem{montiel2015path}
Montiel, O., Orozco-Rosas, U., Sep{\'u}lveda, R.: Path planning for mobile
  robots using bacterial potential field for avoiding static and dynamic
  obstacles.
\newblock Expert Systems with Applications \textbf{42}(12), 5177--5191 (2015)

\bibitem{Nagel}
Nagel, K., Schreckenberg, M.: {A cellular automaton model for freeway traffic}.
\newblock Journal de Physique I \textbf{2}(12), 2221--2229 (1992)

\bibitem{nakagaki2000intelligence}
Nakagaki, T., Yamada, H., T{\'o}th, {\'A}.: Intelligence: Maze-solving by an
  amoeboid organism.
\newblock Nature \textbf{407}(6803), 470 (2000)

\bibitem{Nalpa}
Nalpantidis, L., Sirakoulis, G.C., Gasteratos, A.: Non-probabilistic cellular
  automata-enhanced stereo vision simultaneous localization and mapping.
\newblock Measurement Science and Technology \textbf{22}(11), 114,027 (2011)

\bibitem{nascimento2016multi}
Nascimento, T.P., Concei{\c{c}}ao, A.G., Moreira, A.P.: Multi-robot nonlinear
  model predictive formation control: the obstacle avoidance problem.
\newblock Robotica \textbf{34}(3), 549--567 (2016)

\bibitem{nash2013any}
Nash, A., Koenig, S.: Any-angle path planning.
\newblock AI Magazine \textbf{34}(4), 85--107 (2013)

\bibitem{Neumann}
Neumann, J.V.: Theory of Self-Reproducing Automata.
\newblock University of Illinois Press, Champaign, IL, USA (1966)

\bibitem{nieto2014coordination}
Nieto-Granda, C., Rogers~III, J.G., Christensen, H.I.: Coordination strategies
  for multi-robot exploration and mapping.
\newblock The International Journal of Robotics Research \textbf{33}(4),
  519--533 (2014)

\bibitem{Ntinas}
Ntinas, V.G., Moutafis, B.E., Trunfio, G.A., Sirakoulis, G.C.: Parallel fuzzy
  cellular automata for data-driven simulation of wildfire spreading.
\newblock Journal of Computational Science \textbf{21}(Supplement C), 469 --
  485 (2017)

\bibitem{pandey2014path}
Pandey, A., Sonkar, R.K., Pandey, K.K., Parhi, D.: Path planning navigation of
  mobile robot with obstacles avoidance using fuzzy logic controller.
\newblock In: Intelligent Systems and Control (ISCO), 2014 IEEE 8th
  International Conference on, pp. 39--41. IEEE (2014)

\bibitem{Sun2009}
Q., S., Z.J., D.: A new shortest path algorithm using cellular automata model.
\newblock Computer Technology and Development \textbf{19}(2), 42--44 (2009)

\bibitem{qu2013improved}
Qu, H., Xing, K., Alexander, T.: An improved genetic algorithm with
  co-evolutionary strategy for global path planning of multiple mobile robots.
\newblock Neurocomputing \textbf{120}, 509--517 (2013)

\bibitem{CERBAL}
R., H., V\"olkmann, K.P., Sobolewski, M.: The cellular processing machine cepra
  - 8l.
\newblock Mathematical Research \textbf{81}, 179--199 (1994)

\bibitem{Sirakoulis}
Sirakoulis, G.C., Adamatzky, A.: Robots and Lattice Automata.
\newblock Springer Publishing Company, Incorporated (2014)

\bibitem{acri2012}
Sirakoulis, G.C., Bandini, S. (eds.): Cellular Automata - 10th International
  Conference on Cellular Automata for Research and Industry, {ACRI} 2012,
  Santorini Island, Greece, September 24-27, 2012. Proceedings, \emph{Lecture
  Notes in Computer Science}, vol. 7495. Springer (2012)

\bibitem{Sirakoulis99}
Sirakoulis, G.C., Karafyllidis, I., Sirakoulis, G.C., Mardiris, V.,
  Thanailakis, A.: Study of lithography profiles developed on non-planar si
  surfaces.
\newblock Nanotechnology \textbf{10}(4), 421 (1999)

\bibitem{Sirakoulis99a}
Sirakoulis, G.C., Karafyllidis, I., Soudris, D., Georgoulas, N., Thanailakis,
  A.: A new simulator for the oxidation process in integrated circuit
  fabrication based on cellular automata.
\newblock Modelling and Simulation in Materials Science and Engineering
  \textbf{7}(4), 631 (1999)

\bibitem{stentz1995focussed}
Stentz, A., et~al.: The focussed d\^{}* algorithm for real-time replanning.
\newblock In: IJCAI, vol.~95, pp. 1652--1659 (1995)

\bibitem{syed2014guided}
Syed, U.A., Kunwar, F., Iqbal, M.: Guided autowave pulse coupled neural network
  (gapcnn) based real time path planning and an obstacle avoidance scheme for
  mobile robots.
\newblock Robotics and autonomous systems \textbf{62}(4), 474--486 (2014)

\bibitem{Tsiftsis}
Tsiftsis, A., Georgoudas, I.G., Sirakoulis, G.C.: Real data evaluation of a
  crowd supervising system for stadium evacuation and its hardware
  implementation.
\newblock IEEE Systems Journal \textbf{10}(2), 649--660 (2016)

\bibitem{TsiftsisFPGA}
Tsiftsis, A., Sirakoulis, G.C., Lygouras, J.: {FPGA Processor with GPS for
  Modelling Railway Traffic Flow}.
\newblock Journal of Cellular Automata \textbf{12}(5), 381--400 (2015)

\bibitem{Tsompanas}
Tsompanas, M.A.I., Adamatzky, A., Sirakoulis, G.C., Greenman, J., Ieropoulos,
  I.: Towards implementation of cellular automata in microbial fuel cells.
\newblock PLOS ONE \textbf{12}, 1--16 (2017)

\bibitem{Tsompanas15plus}
Tsompanas, M.A.I., Mayne, R., Sirakoulis, G.C., Adamatzky, A.I.: A cellular
  automata bioinspired algorithm designing data trees in wireless sensor
  networks.
\newblock International Journal of Distributed Sensor Networks \textbf{11}(6),
  471,045 (2015)

\bibitem{Tsompanas2012}
Tsompanas, M.A.I., Sirakoulis, G.C.: Modeling and hardware implementation of an
  amoeba-like cellular automaton.
\newblock Bioinspiration \& Biomimetics \textbf{7}(3), 036,013 (2012)

\bibitem{Tsompanas16}
Tsompanas, M.A.I., Sirakoulis, G.C., Adamatzky, A.: Cellular Automata Models
  Simulating Slime Mould Computing, pp. 563--594.
\newblock Springer International Publishing, Cham (2016)

\bibitem{Tsompanas15}
Tsompanas, M.A.I., Sirakoulis, G.C., Adamatzky, A.I.: Evolving transport
  networks with cellular automata models inspired by slime mould.
\newblock IEEE Transactions on Cybernetics \textbf{45}(9), 1887--1899 (2015)

\bibitem{tsompanas2015slime}
Tsompanas, M.A.I., Sirakoulis, G.C., Adamatzky, A.I.: Evolving transport
  networks with cellular automata models inspired by slime mould.
\newblock IEEE Transactions on Cybernetics \textbf{45}(9), 1887--1899 (2015)

\bibitem{Tsompanas2016}
Tsompanas, M.A.I., Sirakoulis, G.C., Adamatzky, A.I.: Physarum in silicon: the
  greek motorways study.
\newblock Natural Computing \textbf{15}(2), 279--295 (2016)

\bibitem{tzionas1997collision}
Tzionas, P.G., Thanailakis, A., Tsalides, P.G.: Collision-free path planning
  for a diamond-shaped robot using two-dimensional cellular automata.
\newblock IEEE Transactions on Robotics and Automation \textbf{13}(2), 237--250
  (1997)

\bibitem{3D}
Wang, Y.: Study for solving the path on the three-dimensional surface based on
  cellular automata method.
\newblock Modern Applied Science \textbf{4}(5), 196--200 (2010)

\bibitem{Warshall}
Warshall, S.: A theorem on boolean matrices.
\newblock J. ACM \textbf{9}(1), 11--12 (1962)

\bibitem{Was}
Was, J., Sirakoulis, G.C., Bandini, S. (eds.): Cellular Automata - 11th
  International Conference on Cellular Automata for Research and Industry,
  {ACRI} 2014, Krakow, Poland, September 22-25, 2014. Proceedings,
  \emph{Lecture Notes in Computer Science}, vol. 8751. Springer (2014)

\bibitem{WuXue}
X.J., W., H.F., X.: �shortest path algorithm based on cellular automata
  extend model.
\newblock Computer Applications \textbf{24}(5), 92--3 (2004)

\bibitem{Samira}
Yacoubi, S.E., Was, J., Bandini, S. (eds.): Cellular Automata - 12th
  International Conference on Cellular Automata for Research and Industry,
  {ACRI} 2016, Fez, Morocco, September 5-8, 2016. Proceedings, \emph{Lecture
  Notes in Computer Science}, vol. 9863. Springer (2016)

\bibitem{zhang2014rapid}
Zhang, X., Zhang, Y., Zhang, Z., Mahadevan, S., Adamatzky, A., Deng, Y.: Rapid
  physarum algorithm for shortest path problem.
\newblock Applied Soft Computing \textbf{23}, 19--26 (2014)

\end{thebibliography}

\end{document}